\documentclass[aps,pre,showpacs,twocolumn]{revtex4-2}%
\usepackage{graphicx}
\usepackage[T1]{fontenc}
\usepackage{graphicx}
\usepackage{amsmath}
\usepackage{amsfonts}
\usepackage{xcolor}
\providecommand{\LyX}{L\kern-.1667em\lower.25em\hbox{Y}\kern-.125emX\@}

\makeatother
\makeatother
\begin{document}
\title{Stability of rotatory solitary states in Kuramoto networks with inertia}	
\author{Vyacheslav O. Munayev$^{1}$, Maxim I. Bolotov$^{1}$, Lev A. Smirnov$^{1,2}$, Grigory V. Osipov$^{1}$, and Igor V. Belykh$^{3,1}$\footnote{Corresponding author, e-mail: ibelykh@gsu.edu}}
\address{$^1$Department of Control Theory, Lobachevsky State University of Nizhny Novgorod,
	23 Gagarin Avenue, Nizhny Novgorod, 603022, Russia\\
	$^2$Institute of Applied Physics, Russian Academy of Sciences, Ul'yanova Str. 46, Nizhny Novgorod, 603950, Russia\\
	$^3$Department of Mathematics and Statistics, Georgia State University, P.O. Box 4110, Atlanta, Georgia, 30302-410, USA}

\begin{abstract}
Solitary states emerge in oscillator networks when one oscillator separates from the fully synchronized cluster and becomes incoherent with the rest of the network. Such chimera-type  patterns with an incoherent state formed by a single oscillator were observed in various oscillator networks; however, there is still a lack of  understanding of how such states can stably appear. Here, we study the stability of solitary states in Kuramoto networks of identical two-dimensional phase oscillators with inertia and a phase-lagged coupling. The presence of inertia can induce rotatory dynamics of the phase difference between the solitary oscillator and the coherent cluster. We derive asymptotic stability conditions for such a solitary state as a function of inertia, network size, and phase lag that may yield either attractive or repulsive coupling. Counterintuitively, our analysis demonstrates that (i) increasing the size of the coherent cluster can promote the stability of the solitary state in the attractive coupling case and (ii) the solitary state can be stable in small-size networks with all repulsive coupling. We also discuss the implications of our stability analysis for the emergence of  rotatory chimeras.
\end{abstract}
\date{\today}
\maketitle

\section{Introduction}
Networks of phase oscillators are often used to model cooperative dynamics in various biological and man-made systems, ranging from neuronal networks \cite{hoppensteadt2012weakly} and populations of chemical oscillators \cite{tinsley2012chimera} to laser arrays \cite{ding2019dispersive} and power grids \cite{dorfler2013synchronization}. The Kuramoto network of first-order phase oscillators \cite{kuramoto1975self, strogatz2000kuramoto}
is a widely adapted model of a phase oscillator network which 
can exhibit complex spatio-temporal dynamics during the transition from incoherence to full synchronization
\cite{acebron,barreto2008synchronization, ott2008low,hong2007entrainment, pikovsky2008partially, maistrenko2004mechanism,dorfler2011critical}. When oscillators in the Kuramoto model have heterogeneous frequencies, this transition is typically accompanied by partial synchronization which emerges  
when the network splits into clusters of coherent and incoherent oscillators 
 \cite{acebron,martens2009exact,laing2009dynamics}. In the case of identical oscillators, partial synchronization can turn into chimera states that represent fascinating patterns
in which even structurally identical oscillators can break into two, possibly asymmetric groups of coherent and incoherent oscillators \cite{kuramoto2002coexistence, abrams2004chimera, abrams2008solvable, panaggio2015chimera}. Chimera states were extensively studied in the Kuramoto model as well as in other networks of oscillatory  systems \cite{omel2013coherence, wolfrum2011spectral,laing2019dynamics,ashwin2015weak,panaggio2015chimera,panaggio2016chimera,larger2013virtual,semenova2016coherence,bolotov2016marginal,bolotov2018simple}, including coupled chemical oscillators \cite{tinsley2012chimera}, networks of metronomes \cite{martens}, coupled pendula \cite{kapitaniak2014imperfect}, pedestrians on a bridge \cite{belykh2017foot}, optical systems and lasers \cite{hagerstrom2012experimental}, and continuous media \cite{nicolaou2017chimera,smirnov2017chimera}. 

The second-order Kuramoto model with inertia is commonly used for describing networks of oscillators capable of adjusting their natural frequencies as, for example, in the adaptive frequency model of firefly synchronization \cite{ermentrout1991adaptive} and power grid systems \cite{tumash2019stability}. The inclusion of inertia yields two-dimensional intrinsic oscillator dynamics, thereby making the cooperative dynamics of the second-order Kuramoto model substantially more complex than of its classical first-order counterpart. These inertia-induced dynamics include  complex and hysteretic transitions from incoherence to full synchronization  \cite{tanaka1997first,tanaka1997self,ji2014low,munyaev2020analytical,komarov2014synchronization,olmi2014hysteretic,barabash2021partial}, bistability of synchronous clusters \cite{belykh2016bistability}, chaotic inter-cluster dynamics \cite{brister2020three}, 
chimeras \cite{olmi2015chimera,maistrenko2017smallest,medvedev2021stability}, and 
solitary states \cite{jaros2015chimera,jaros2018solitary}. Solitary states emerge in a network 
at the edge of full synchronization when all but one oscillator synchronize in a synchronous cluster, while the remaining, ``solitary'' oscillator has either a constant phase shift with respect to the synchronous cluster or rotates with a different frequency. The emergence of such solitary states in networks of identical second-order Kuramoto oscillators was numerically studied in \cite{jaros2018solitary,maistrenko2017smallest}. In particular, Maistrenko et al.  \cite{maistrenko2017smallest} showed that solitary states, termed as weak chimeras which are formed by a two-oscillator synchronous cluster and one incoherent oscillator can appear even in smallest, three-node networks of Kuramoto oscillators with inertia. Alternatively, solitary states can be viewed as a particular case of multi-cluster synchronization whose existence  and stability were analyzed in two- and three-population Kuramoto networks \cite{belykh2016bistability,brister2020three}.  More precisely, Belykh et al. \cite{belykh2016bistability}  analytically studied the emergence and co-existence of stable clusters in a two-population network of identical second-order Kuramoto oscillators. The two populations of different sizes $N$ and $M$ could naturally split into two clusters where the oscillators synchronize within a cluster while creating a phase shift  between the clusters. 
The dynamics of this phase shift is governed by a driven pendulum equation, and therefore the phase shift can be constant or can vary from $0$ to $2\pi$, inducing a two-cluster pattern with a rotatory phase shift. The analysis performed in  \cite{belykh2016bistability}  yielded  necessary and sufficient conditions for the stability of two-cluster synchronization with a constant phase and also provided a proof-of-concept stability condition for  two-cluster synchronization with a rotating phase shift. These stability results are directly 
applicable to solitary states in the two-population network with one population composed of a single oscillator $M=1,$ under the constraint that the intra-coupling within each population is not equal to the inter-coupling between the populations. 

In this paper, we seek to relax this constraint and derive stability conditions
for solitary states in the second-order Kuramoto model of identical oscillators with  homogeneous, phase-lagged Kuramoto-Sakaguchi coupling. In this setting, such solitary states may only have a stable rotatory phase shift between the solitary oscillator and the rest of the network. 
Apart from the
conservative proof-of-concept condition \cite{belykh2016bistability}, there is a lack of  analytical results on the stability of rotatory solitary states. Here, we aim to 
close this gap by performing an asymptotic stability analysis of solitary states and providing approximate stability bounds that relate inertia, network size, and phase lag. The phase lag is allowed to change from $0$ to $\pi,$ so that the coupling may be attractive or repulsive.
In particular, our analysis predicts that increasing the network size, and therefore the size of the coherent cluster in a network with all attractive coupling, can promote the stability of a rotatory solitary state. We find this dependence  counterintuitive as one would expect that such an increase would make the ``mass'' of the coherent cluster greater, thereby enhancing its ``gravitation force'' that would attract the solitary oscillator back to the cluster. Instead, the opposite happens. Through analysis and numerics, we also demonstrate that rotatory solitary states can be stable in small-size networks with all repulsive coupling.

The emergence of solitary states as a result of competing attractive and repulsive coupling
in the first-order Kuramoto model with mixed coupling was previously studied in the cases of identical \cite{maistrenko2014solitary} and non-identical oscillators \cite{teichmann2019solitary}. Our analysis of rotatory solitary states in the second-order Kuramoto model that are impossible in its first-order counterpart points out to a different role of purely repulsive coupling that can act similarly to common inhibition in networks of bursting neurons \cite{belykh2008weak}. More precisely, global repulsive coupling plays the role of common inhibition that allows the solitary oscillator to synchronize the other oscillators into the synchronous cluster, while maintaining its phase independence via a delicate balance of common and mutually inhibitory forces. 

The layout of this paper is as follows. In Sec.~II, we introduce the oscillator network model and state the problem under consideration. In Sec. III, we study the existence of solitary states and show that the dynamics of the phase difference between the solitary oscillator and synchronous cluster is governed by the classical pendulum equation with damping and constant torque. This dynamics can induce a rotatory solitary state. 
In Sec. IV, we analyze the variational equations for the stability of the rotatory solitary state and obtain the main stability results of this paper. We also numerically validate our theoretical results. In Sec.~V, we numerically analyze attraction basins of solitary networks  states in different size networks and show that rotatory solitary states are resilient patterns that robustly appear from a large set of randomly chosen initial conditions.  Section~VI contains concluding remarks and discussions. Appendices A and B contain derivations of the stability conditions.

\section{Network model and solitary states}
We consider a globally coupled network of two-dimensional Kuramoto oscillators with inertia:
\begin{equation}
m\ddot{\theta}_i+\dot{\theta}_i=\omega+\frac{1}{N}\sum\limits_{\tilde{j}=1}^{N}\sin{\left(\theta_{j}-\theta_i-\alpha\right)},
\label{eq:mainSys}
\end{equation}
where variables $\theta_i,$  $i=1,...,N$ represent the oscillators' phases. The oscillators are assumed to be identical, with identical frequency $\omega,$ inertia $m,$ and phase lag $\alpha \in [0, \pi)$ that represents the Kuramoto-Sakaguchi coupling \cite{sakaguchi2006instability}.
System \eqref{eq:mainSys} has complete synchronization manifold $D(1)=\{\theta_1=...=\theta_N\}$ which is locally stable for any $\alpha \in [0, \pi/2)$ and 
unstable for any $\alpha \in (\pi/2,\pi)$ \cite{acebron}. As a result, the Kuramoto-Sakaguchi coupling is termed as {\it attractive} for $\alpha<\pi/2$ and {\it repulsive} for $\pi/2<\alpha<\pi.$ Due to the all-to-all symmetric coupling, the network decomposition supports the complete set of all possible clusters. These clusters are represented by disjoint groups of oscillators defined by the equalities of the oscillator states. In this paper, we will focus on the dynamics and stability of a particular two-cluster partition that corresponds to a solitary state and is determined by the manifold 
\begin{equation}
		D(2)=	\left \{	\begin{array}{l}
	\theta_1=\Theta_s,\dot{\theta}_1=\dot{\Theta}_s,\\
		\theta_2=...=\theta_N=\Theta,
		\dot{\theta}_2=...=\dot{\theta}_N=\dot{\Theta}
	\end{array}\right\}.
	\label{cluster}
\end{equation}
This cluster pattern is termed as a solitary state in which a single non-synchronized oscillator forms the first cluster while the remaining oscillators compose the second synchronized cluster. The phase difference between the first and second clusters may  be time-dependent and periodically vary from $0$ to $2\pi.$ The choice of the first oscillator as the solitary oscillator in $D(2)$ is arbitrary, and due to the symmetry, there are other $N-1$ solitary states that have identical dynamical and stability characteristics. The emergence of a particular solitary state depends on the choice of initial conditions.

The dynamics on solitary state manifold $D(2)$ is governed by the four-dimensional system 
\begin{equation}
	\begin{gathered}
		m\ddot{\Theta}_s+\dot{\Theta}$  $_s=\omega-\frac{1}{N}\sin{\alpha}+\frac{N-1}{N}\sin{\left(\Theta-\Theta_s-\alpha\right)},\\
		m\ddot{\Theta}+\dot{\Theta}=\omega-\frac{N-1}{N}\sin{\alpha}+\frac{1}{N}\sin{\left(\Theta_s-\Theta-\alpha\right)}.
	\end{gathered}\label{manifold}
\end{equation}
We seek to characterize possible dynamics of system \eqref{manifold} and 
derive approximate conditions for transversal stability of solitary state manifold $D(2),$ thereby revealing the role of inertia, the network size, and the phase lag in the emergence of stable solitary states.

\section{Existence of rotatory solitary states}\label{existence}
Introducing the difference between the phases of the solitary oscillator and synchronous cluster, 
$X=\Theta_s-\Theta,$ and subtracting the second equation from the first equation in \eqref{manifold}, we obtain
\begin{equation}
	\begin{array}{rcl}
m\ddot{X}+\dot{X}&=&  \frac{1}{N} [(N-2) \sin{\alpha}-\\
&-&\{ \left (N-1)\sin{\left(X+\alpha\right)}
+ \sin{\left(X-\alpha\right)}\right\}].
\end{array}
\label{eq:detuning}
\end{equation}
\subsection{Transformation to the pendulum equation}
As in  \cite{belykh2016bistability}, we simplify the expression in the curly brackets in \eqref{eq:detuning} by means of the trigonometric identity
$$	\begin{array}{l}
	(N-1)\sin{\left(X+\alpha\right)}
	+ \sin{\left(X-\alpha\right)}=\\
(N\cos \alpha \sin X+(N-2)\sin \alpha \cos X)=R\sin(X+\delta),
	\end{array}
$$
where 
\begin{equation}
R=\sqrt{\left(N-1\right)^2+1+2\left(N-1\right)\cos{2\alpha}}
\label{R}
\end{equation}
 and
$\delta=\arccos (\frac{N}{R}\cos{\alpha}).$ We then rescale time 
$t=\dfrac{N}{\rho R}\hat{t},$ where $\hat{t}$ is a new time and $\rho=\sqrt{\frac{N}{mR}}$ and use substitution $X+\delta=\Phi$ to turn \eqref{eq:detuning} into the pendulum equation 
\begin{equation}
	\dfrac{d^2\Phi}{d\hat{t}^2}+ \rho \dfrac{d\Phi}{d\hat{t}}+\sin\Phi =\gamma,\label{pendulum}
\end{equation}
where $\gamma=\frac{N-2}{R}\sin{\alpha}$ represents constant torque. Pendulum equation \eqref{pendulum} is well-known to exhibit oscillatory dynamics, determined by a stable rotatory limit cycle that may co-exist with a stable fixed point \cite{andronov2013theory}. As a result, the phase intercluster difference between the solitary state and synchronous cluster, whose dynamics is governed by \eqref{pendulum}, can periodically oscillate, while making complete turns around
the cylinder $(\Phi\;{\rm mod} 2\pi,$ $\dot{\Phi}=v)$ or can be constant. The relation between pendulum equation \eqref{pendulum} and intercluster dynamics of the second-order two-population Kuramoto system was studied in detail for  $\alpha \in [0, \pi/2)$ in \cite{belykh2016bistability}. Here, we briefly discuss subtle differences imposed by the homogeneous coupling in the one-population network \eqref{eq:mainSys} and  $\alpha \in [\pi/2, \pi).$\\
{\it Case 1. Attractive coupling with $\alpha \in [0, \pi/2)$.}
The well-known stability diagram \cite{andronov2013theory,belykh2016bistability} for pendulum equation \eqref{pendulum} contains three regions of distinct dynamics as a function of damping parameter $\rho$ and constant torque $\gamma.$ These regions are separated by homoclinic bifurcation curve $\gamma=T(\rho),$ often called the Tricomi curve \cite{tricomi1933integrazione}, and saddle-node bifurcation line $\gamma=1.$ 
While there is no exact equation for the homoclinic bifurcation curve, it can be approximated rather precisely \cite{belykh2016bistability} as
\begin{equation}
	\gamma=T(\rho)=\frac{4}{\pi}\rho- 0.305 \rho^3.
	\label{tricomi}
\end{equation}
In Region~1 under curve \eqref{tricomi} (see the convention used in Fig.~1 in  \cite{belykh2016bistability}), pendulum system (\ref{pendulum}) has two fixed points on the cylinder: a stable fixed point $\Phi_e=\arcsin \gamma$ and a saddle $\Phi_s=\pi-\arcsin \gamma.$ In Region~2, bounded by the homoclinic bifurcation curve and the saddle-node bifurcation line, these fixed points co-exist with a rotatory stable limit cycle $\Phi_c(t),$ emerged  from a homoclinic orbit of saddle $\Phi_s$ at $\gamma=T(\rho).$ Finally, in Region~3, system (\ref{pendulum}) only has the limit cycle  $\Phi_c(t).$ 

\begin{figure}[h!]\center
	\includegraphics[width=8.6cm]{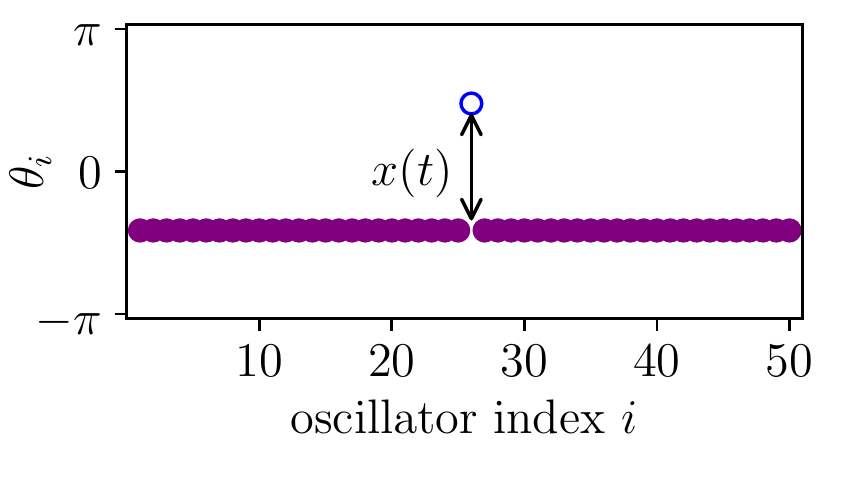}
	\caption{Snapshot of a rotatory solitary state in network \eqref{eq:mainSys} of 50 oscillators. Time-varying $x(t)$ governs the phase difference between the solitary oscillator (blue open circle) and the synchronous cluster (purple circles). Parameters are $\omega=1,$ $m=20,$ and $\alpha=\pi/6.$
	}\label{snapshot}
\end{figure}

Note that stable fixed point $\Phi_e$ corresponds to phase difference $X_e=\Phi_e-\delta=0$
in the original system \eqref{eq:detuning}. In contrast to the two-population setting in \cite{belykh2016bistability} with a non-zero
constant phase shift $X_e$, system \eqref{eq:detuning} yields the zero constant phase difference so that the corresponding solitary state simply turns into complete synchronization. Therefore, network \eqref{eq:mainSys} may only have stable rotatory solitary states with non-zero phase difference $X(t),$ governed by stable limit cycle $x(t)\equiv X_c(t)=\Phi_c(\hat{t})-\delta.$ Figure~\ref{snapshot} illustrates the existence of such a solitary state with rotatory phase difference $x(t).$\\
{\it Case 2. Repulsive coupling with $\alpha \in (\pi/2, \pi)$.}
In this case, the dynamics of pendulum equation \eqref{pendulum} is essentially the same as for $\alpha \in [0, \pi/2),$ except for an important caveat that fixed points  $\Phi_e=\arcsin \gamma$ and  $\Phi_s=\pi-\arcsin \gamma$ exchange their roles and become a saddle and a stable fixed point, respectively. Therefore, the constant non-zero phase shift, defined by 
$\Phi_s=\pi-\arcsin \gamma,$ yields a solitary state with a constant phase difference. However, this solitary state is always unstable (see the next section for the proof), so that network \eqref{eq:mainSys} may only have stable rotatory solitary states for all  $\alpha \in (0, \pi)$.  Regions 1-3 for the repulsive coupling are bounded by curves \eqref{tricomi} with negative $\gamma$ and curve $\gamma=-1.$ 

Therefore, in both attractive and repulsive coupling cases, the existence region for rotatory solitary states is determined by condition $\gamma \ge T(\rho).$
In terms of difference system \eqref{eq:detuning} and the original parameters of system \eqref{eq:mainSys}, this condition takes the form
\begin{equation}
	\dfrac{N-2}{R}\left|\sin{\alpha}\right|\ge T\left(\sqrt{\dfrac{N}{mR}}\right),\label{bound}
\end{equation}
where $T\left(\cdot\right)$ and $R$ are given in \eqref{tricomi} and \eqref{R}, respectively.
The equality sign in \eqref{bound} yields the bound for the existence of rotatory solitary states (the black lines in Fig.~\ref{main}): 
\begin{equation}
	m\left(\alpha\right)=\dfrac{N}{R
	\left[T^{-1}\left(\left|\sin{\alpha}\right|\right)\right]^{2}},\label{m1}
\end{equation}
where $T^{-1}$ is the inverse function of $T.$ It follows from \eqref{m1} that increasing  network size $N$ also increases the size of the existence region for a rotatory solitary state (see Fig.~\ref{main} ).
\subsection{Analytical estimates for the rotatory limit cycle}
The solution for rotatory limit cycle $x(t)$ which governs the dynamics of the intercluster phase difference cannot be written in close form. When the parameters of pendulum system \eqref{pendulum} are chosen slightly above the homoclinic bifurcation curve $T(\rho),$
the rotatory limit cycle $x(t)$ inherits the shape of the homoclinic orbit of saddle $\Phi_s$ for $\alpha \in (0,\pi/2)$ ($\Phi_e$ for $\alpha \in (\pi/2,\pi)$). As a result, the limit cycle spends most of the time in the vicinity of the saddle while making fast excursions around the cylinder, and therefore, $x(t)$ can be approximated by the coordinate of the saddle. This conservative, proof-of-concept bound was used in \cite{belykh2016bistability,brister2020three} to analyze the stability of inter-cluster rotatory dynamics. In the following, we will derive less conservative analytical estimates for rotatory solution $x(t)$ which are applicable to much broader parameter regions, thereby enabling a general analytic stability analysis of rotatory solitary states.

We apply the Lindstedt--Poincar\'e method \cite{drazin1992nonlinear} to approximate the rotatory, periodic solution $x(t)$ in the case of large inertia, $m \gg 1.$ The method introduces a new scaled time and seeks the solution and the time scaling as an asymptotic series, thereby removing secular terms - terms that grow without bounds. This approximate solution derived in the limit of large inertia has the form
\begin{equation}
	\begin{array}{rcl}
		x\left(t\right)&=&\omega_x t+\dfrac{\varepsilon^2}{\Omega_p}\left(\cos\omega_x t-1+\Omega_p^{-1}\cos\alpha\sin\omega_x t\right)\\
		&+&\dfrac{\varepsilon^4}{4\Omega_p^3}\!\left(\!\cos\alpha\!\left(\cos2\omega_x t-1\right)\!+\!\dfrac{\cos^2\!\alpha\!-\!\Omega_p^2}{2\Omega_p}\sin2\omega_x t\right)\\
		&+&\ldots+o\left(\varepsilon^8\right), \\
		\omega_x&=&\Omega_p\!-\!\dfrac{\varepsilon^4}{2N^2\Omega_p^3}\!\left(\left(N\!-\!1\right)^2+1+2\left(N\!-\!1\right)\cos2\alpha\right) \\
		&+&\ldots+o\left(\varepsilon^8\right),
	\end{array}
	\label{approx}
\end{equation}
where $\Omega_p=\frac{N-2}{N}\sin\alpha$ and $\varepsilon=1/\sqrt{m}\ll 1$ is a small parameter.
The complete solution with higher-order terms corresponding to coefficients  $\varepsilon^3,..,\varepsilon^8$ along with its derivation are given in Appendix A. The inclusion of the terms up to the eight power is necessary for the stability conditions using \eqref{approx} to adequately describe the stability boundary (the dashed pink line in Fig.~\ref{main}) and its curvature at the values of $\alpha$ close to $\pi/2.$ This point will be discussed in more detail in the next section. We will also demonstrate that the predictive power of approximate solution \eqref{approx} extends far beyond the large inertia case and yields remarkably close stability bounds for relatively small inertia $m>5.$

\section{Stability analysis}\label{stability}
To derive analytical stability conditions for the emergence of solitary states in the network, we seek to prove the local transversal stability of solitary state manifold $D(2).$ 
To do so, we first introduce infinitesimal difference variables for the oscillators from the synchronous cluster:
\begin{equation}
	\begin{array}{l}
		\eta_{k}=\eta_k-\eta_{k+1},\;\;k=2,...,N-1.
	\end{array} \label{eta}
\end{equation}
We then subtract the $(k+1)$-th equation from the $k$-th equation in system \eqref{eq:mainSys} and by virtue of \eqref{eta} obtain the following variational equations for the transversal stability:
\begin{equation}
	\begin{array}{l}
		m\ddot{\xi}_k+\dot{\xi}_k+ \left(\dfrac{N-1}{N}\cos{\alpha}+\dfrac{1}{N}\cos{\left(X-\alpha\right)}\right)\xi_k=0,
	\end{array}
	\label{eq:stability_xi}
\end{equation}
where $k=\overline{2,N-1}$ and phase difference $X$ between the solitary state and the synchronous cluster is governed by system \eqref{eq:detuning}. The stability of this linearized 
second-order differential equation with a possibly time-varying coefficient via $X(t)$ implies the local transversal stability of $D(2).$ Note that  \eqref{eq:stability_xi} is a system of uncoupled identical equations such that the stability of one equation implies the stability of all the others.  Therefore, hereafter we will only investigate the stability of one of the $k$-th equations, omitting subscript $k.$

\subsection{Instability of solitary states with a constant phase difference}
As shown in Sec.~\ref{existence}, constant phase difference $X$ is non-zero only for  $\alpha\in(\pi/2,\pi)$ so that 
$X=X_s=2\arctan\!\left(\frac{N}{N-2}\cot\alpha\right)$. Substituting $X=X_s$ into variational equation \eqref{eq:stability_xi} and removing subscript $k$, we obtain the linear differential equation with constant coefficients: 
\begin{equation}
	\begin{array}{l}
		m\ddot{\xi}+\dot{\xi}+ \left(\dfrac{N-1}{N}\cos{\alpha}+\dfrac{1}{N}\cos{\left(X_s-\alpha\right)}\right)\xi=0.
	\end{array}
	\label{constant}
\end{equation}
Its characteristic equation takes the form
\begin{equation}
	s^2+\frac{1}{m}s+\frac{1}{m}\frac{\left(N-2\right)N\cos\alpha}{\left(N-2\right)^2\sin^2\alpha+N^2\cos\alpha ^2}=0.\label{s}
\end{equation}
Note that the third term in \eqref{s} is always negative in the considered range of $\alpha\in(\pi/2,\pi)$ since $\cos \alpha<0.$ Therefore, one of the roots $s_{1,2}$ is positive so that linear differential equation \eqref{constant} is unstable. Thus, the solitary state with a constant phase shift is always unstable, and only rotatory solitary states can stably emerge in network \eqref{eq:mainSys}. 

\subsection{Stability of rotatory solitary states}
Time-varying dynamics of phase shift $X(t)$ governed by rotatory limit cycle $x(t)$ transform the  variational equation \eqref{constant} into
\begin{equation}
	\begin{array}{l}
		m\ddot{\xi}+\dot{\xi}+ \left(\dfrac{N-1}{N}\cos{\alpha}+\dfrac{1}{N}\cos{\left(x(t)-\alpha\right)}\right)\xi=0.
	\end{array}
	\label{x}
\end{equation}
Due to the presence of the time-varying coefficient, a complete stability analysis of variational equations
\eqref{x} for the full range of parameters seems to be out of reach. Instead, we seek to quantify all possible scenarios of emergent instability in \eqref{x} and derive approximate explicit bounds for the loss of stability in the case of large inertia $m$ and study their applicability to smaller $m.$

Note that the trace of the monodromy matrix of equation \eqref{x} 
\begin{equation}
	A=\begin{pmatrix}
		0 & 1\\
		-\frac{1}{m}\left(\frac{N - 1}{N}\cos\alpha+\frac{1}{N}\cos \left(x\left(t\right)-\alpha\right)\right) & -\frac{1}{m}
	\end{pmatrix}
\end{equation}
is ${\rm trace} A=-1/m.$ Therefore, by virtue of Liouville's formula, we can interconnect  
the eigenvalues of $A$ (Floquet characteristic  multipliers $\mu_1$ and $\mu_2$) via identity $\mu_1\mu_2=\exp{\left(-T_x/m\right)},$ where $T_x$ is the period of limit cycle $x(t).$
In general, there are three possible bifurcation scenarios by which the trivial solution of linear variational equation \eqref{x} can lose its stability. In the case of real multiplies $\mu_1$ and $\mu_2$, these scenarios are associated with (i) a bifurcation when one of the multiplies equals $+1$ and (ii) a bifurcation when one of the multiplies equals $-1.$ In the case of complex multiplies, the instability emerges when the complex multipliers cross the unit circle. However, due to the constraint $\left|\mu_{1,2}\right|^2=\exp{\left(-T_x/m\right)}<1$ for $m>0$, the complex multipliers of matrix $A$ are always located inside the unit circle, and therefore, this third bifurcation transition to instability is impossible for system \eqref{x}. Thus, we shall find the conditions under which system \eqref{x} is close to multiplier $+1$ and $-1$ bifurcations, thereby deriving bounds for the system's instability explicit in the parameters of network \eqref{eq:mainSys}. Note that  multiplier $+1$ and $-1$ bifurcations
in linearized system  \eqref{x} differ from their corresponding counterparts of a nonlinear system such as pitchfork and  period-doubling bifurcations which yield additional periodic solutions in a nonlinear system. On the contrary,  the multiplier $+1$ and $-1$ bifurcations
in the linearized system may only change the stability of the trivial solution, transforming it into a saddle.\\
{\it Scenario A.}
We first aim to find an approximate bifurcation curve associated with stability loss via the  multiplier  $+1$ bifurcation. Our analysis leads to the following assertion.\\
{\bf Statement 1.} [Stability loss near the transition from attractive to repulsive coupling]. \\
{\it Solitary state $D(2)$ with rotatory phase difference $x(t)$ 
	loses its local transversal stability for large inertia $m \gg 1$ at a critical value of phase lag $\alpha$ which is approximated by 
	\begin{equation}
		\begin{array}{l}
		\alpha_c=\dfrac{\pi}{2}+\dfrac{N}{2m\left(N-2\right)^2}-\dfrac{N^3\left(18N^2-27N+13\right)}{96m^3\left(N-2\right)^6}+\\
		~~~~~~ +o\left(m^{-3}\right).
		\end{array}
		\label{eq:stability_right_border}
	\end{equation}
}
{\em Proof}. The proof of Statement~1 is based on an asymptotic analysis of characteristic exponents 
$\Lambda$ and $\hat{\Lambda},$ associated with multipliers $\mu_1$ and $\mu_2$ of equation \eqref{x}, respectively.
Similarly to the analysis performed in Appendix~A for finding an approximate solution \eqref{approx} for rotatory limit cycle $x(t)$ in pendulum-type equation \eqref{eq:detuning}, we seek and approximate  solution $\xi\left(t\right)=e^{\Lambda t}\zeta\left(t\right)$ of variational equation \eqref{x} via expanding both $2\pi$-periodic function $\zeta$ and $\Lambda$ as power series of small parameter $\varepsilon=1/\sqrt{m}.$ Taking into account the terms up to the eight-order of $\varepsilon$, we obtain an approximate expression for characteristic exponent 
$\Lambda.$ Setting $\Lambda=0$ corresponding to multiplier $\mu_1=+1,$  and solving this equality for $\alpha,$ we obtain the approximate bound  \eqref{eq:stability_right_border}. The details of this derivation are delegated to Appendix B.  $\Box$\\
{\it Remark~1.} The proof given in Appendix~B also indicates that characteristic exponent 
$\Lambda$ is negative (positive) for $\alpha<\alpha_c$ ($\alpha>\alpha_c)$ in the vicinity of $\alpha_c,$ thereby analytically demonstrating that increasing $\alpha$ near $\pi$ destabilizes the rotatory solitary state.

\begin{figure*}\center
	\includegraphics[width=1.9\columnwidth]{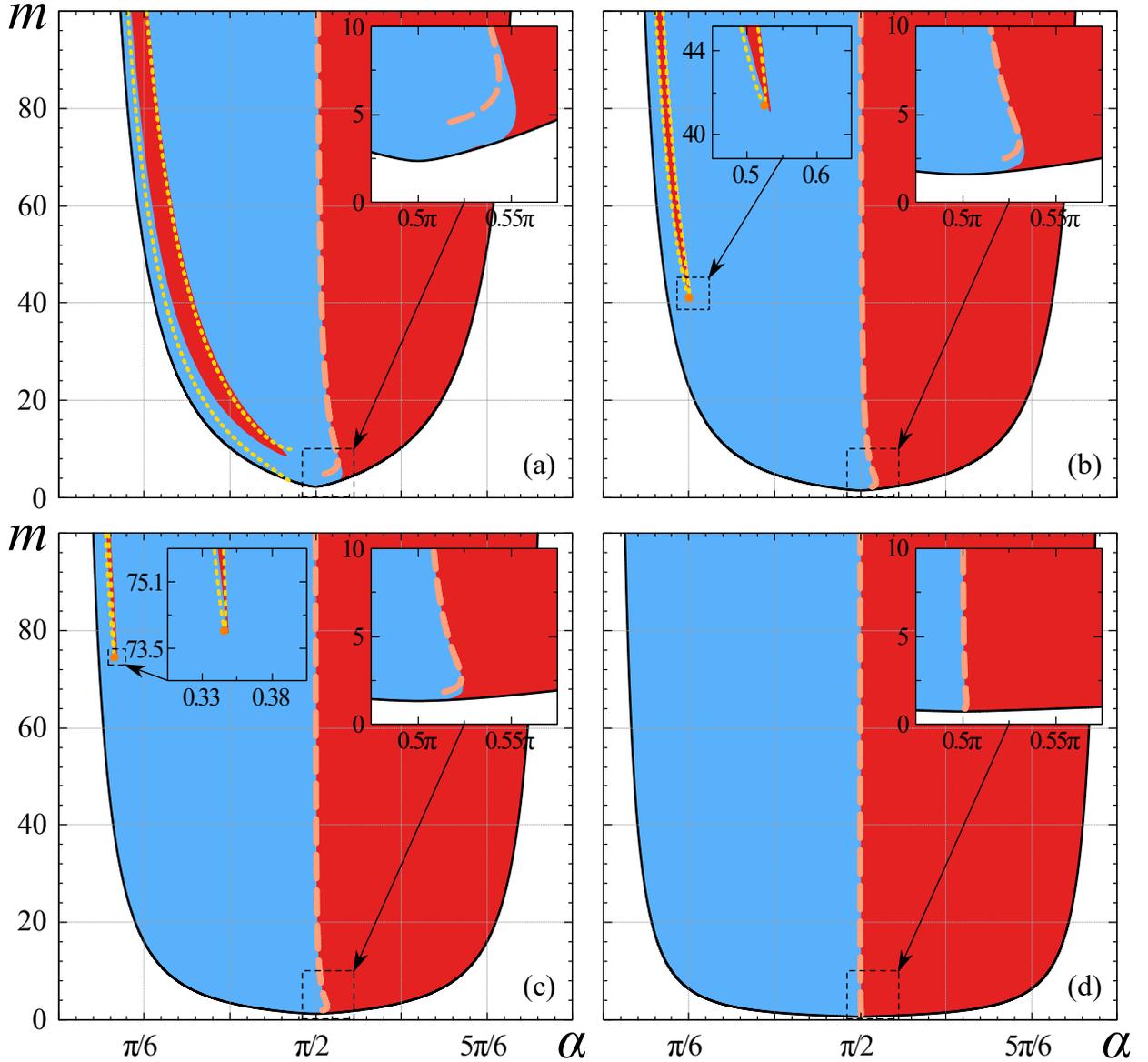}
	\caption{Existence and stability diagrams for the rotatory solitary state. The black solid line depicts the existence boundary defined by homoclinic bifurcation curve \eqref{m1}. The blue (red) region corresponds to local stability (instability) of the rotatory solitary state, evaluated via numerically calculated characteristic exponents of variational equation \eqref{x}. The dashed pink line corresponds to analytical curve \eqref{eq:stability_right_border}. The dotted orange lines are analytical boundaries \eqref{resonance} for the parametric resonance region (red crescent-shaped area). The orange point whose vicinity is zoomed-in in the left insets of panels (b) and (c) indicates the prediction of the
		lowest point of the parametric resonance region via \eqref{eq:parametric_location}. The upper right insets
		detail the shape of analytical curve \eqref{eq:stability_right_border} in the region around $\alpha=\pi/2$ and small $m.$ Notice the stability region for $\alpha>\pi/2$ where the coupling is repulsive. 
		Panel (a): $N=3.$ Panel (b): $N=4$. Panel (c): $N=5.$ Panel (d): $N=50$. Parameter $\omega=1.$}
	\label{main}
\end{figure*}

\begin{figure}[h!]
	{\large (a)}~~~~~~~~~~~~~~~~~~~~~~~~~~~~~~~~~~~~~~~~~~~~~~~~~~~~~~~~~~~~~~~~~~~\\
	\includegraphics[width=0.9\columnwidth]{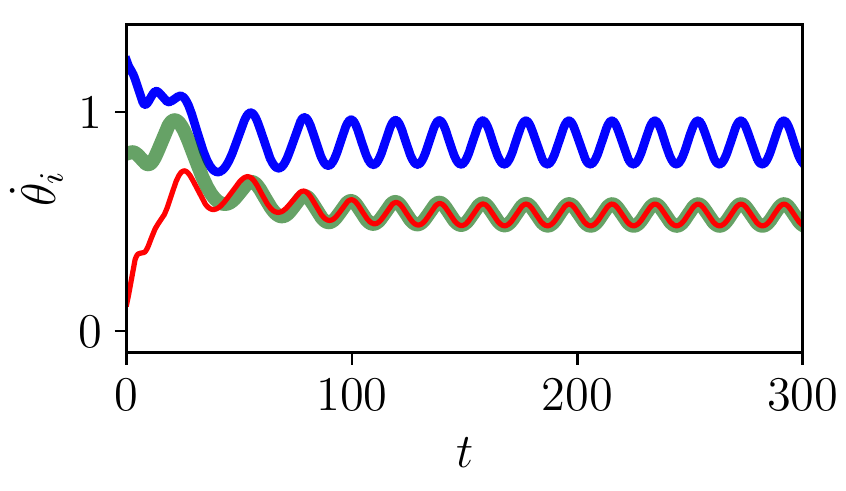}\\
	{\large (b)}~~~~~~~~~~~~~~~~~~~~~~~~~~~~~~~~~~~~~~~~~~~~~~~~~~~~~~~~~~~~~~~~~~~\\
	\includegraphics[width=0.9\columnwidth]{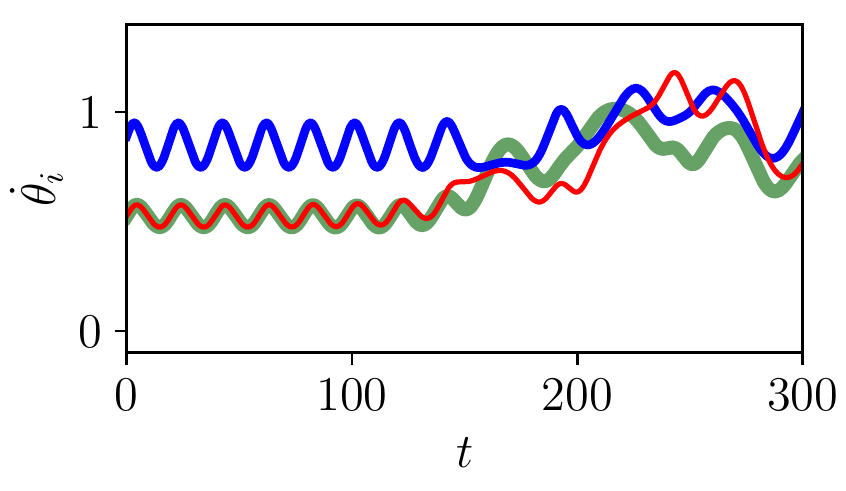}
	\caption{(a) Time series of a stable rotatory solitary state induced by weakly repulsive coupling for $\alpha=\pi/2+0.02$. The established phase difference between the solitary oscillator (blue) and two synchronized oscillators (red and green) is not constant but oscillates with a small amplitude. Initial 
		conditions $\theta_i(0)$ and $\dot{\theta}_i(0)$ are evenly distributed within $[-\pi, \pi]$ and $[0, 2]$, respectively. (b) Increased $\alpha=\pi/2+0.2,$ making the repulsive force stronger,  destabilizes the solitary state. Other parameters $N=3$, $m=20$, $\omega=1.2$.}\label{time-series}
\end{figure}

Figure~\ref{main} indicates that bound \eqref{eq:stability_right_border} (the dashed pink line) predicts the actual change of stability around $\alpha=\pi/2$ remarkably well even for intermediate values of inertia $m$. The predictive power of \eqref{eq:stability_right_border} improves significantly with increasing the network size from inertia $m>8$ in the smallest network with $N=3$ to $m>2$ for $N=5$ and to the minimum  value of $m$ that guarantees the existence of the rotatory solitary state for $N=50$ (compare the insets in Fig.~\ref{main}). The bound \eqref{eq:stability_right_border} also predicts the stability of the rotatory solitary state in a small region of $\alpha>\pi/2,$ corresponding to the repulsive coupling (see the insets in Figs.~\ref{main}(a)-(c) for $N=3,4,5$). Although, it is expected that 
the repulsive coupling promotes desynchronization between the solitary oscillator and the synchronous cluster, we find it surprising that this coupling also maintains synchronization within the synchronous cluster. We hypothesize that this effect might be similar to synchronization in inhibitory networks of bursting neurons in which  a common (possibly, weak) inhibitory drive from a single neuron (representing the solitary oscillator) can synchronize a group of neurons despite their mutual (possibly, strong) repulsive inhibitory connections, provided that a balance between the relative phases of the coupling forces is maintained  \cite{belykh2008weak,shilnikov2008polyrhythmic}. Figure~\ref{time-series}(a) displays the onset of a rotatory solitary state in three-oscillator network \eqref{eq:mainSys} with the parameters from the stability region of the inset in Fig.~\ref{main}(a) with $\alpha>\pi/2,$ yielding  (weakly) repulsive coupling. Increasing  $\alpha$ strengthens the repulsive coupling and destroys the solitary state (Fig.~\ref{time-series}(b)).

{\it Scenario B.} To relate the multiplier $-1$ bifurcation to a resonant ratio of the natural frequency of system \eqref{x} and the frequency of its driving force $x(t),$ we transform \eqref{x} into 
\begin{equation}
	\ddot{\xi}+\frac{1}{m}\dot{\xi}+\hat{\Omega}^2\left(1+q\cos{\left(x\left(t\right)-\alpha\right)}\right)\xi=0,
	\label{eq:Mathieu}
\end{equation}
where $\hat{\Omega}=\sqrt{\dfrac{N-1}{mN}\cos{\alpha}}$ and $q=\dfrac{1}{\left(N-1\right)\cos\alpha}$. Note that equation \eqref{eq:Mathieu} represents a variation of the damped Mathieu equation (also known as a parametric oscillator) that exhibits parametric resonance when time-dependent parameters vary at roughly twice the natural oscillator frequency \cite{taylor1969stability}. Therefore, we expect variational system   \eqref{eq:Mathieu} to become unstable via parametric resonance when the frequency of limit cycle $x(t),$ 
$\omega_x,$ defined in \eqref{approx}, approximately matches the double natural frequency of system \eqref{eq:Mathieu} for small $q$, i.e. when 
\begin{equation}
\omega_x\approx2\hat{\Omega}\;\mbox{\rm  for}\; N \gg 1 .
	\label{double}
\end{equation}
This parametric instability emerging at the double frequency of perturbations of the solitary state can be viewed as an analog of instabilities caused by a period-doubling bifurcation in nonlinear systems, associated with a multiplier -1 bifurcation.  Analyzing solutions of \eqref{eq:Mathieu} under condition \eqref{double}, we derive the following instability criterion.\\
{\bf Statement 2.} [Instability via parametric resonance]. \\
{\it A. Solitary state $D(2)$ with rotatory phase difference $x(t)$ 
	becomes transversely unstable in a region of parameters $(\alpha,m)$ whose boundary can be approximated under the assumption of $m,N \gg 1$ via the implicit function
	\begin{equation}
		\begin{gathered}
			\left(\sqrt{\frac{\left(N-1\right)\cos\alpha}{mN}}-\frac{\left(N-2\right)\sin\alpha}{2N}\right)^2\\
			=\frac{1}{16mN\left(N-1\right)\cos\alpha}-\frac{1}{4m^2}.
		\end{gathered}
		\label{resonance}
	\end{equation}\\
	B. The minimum value of inertia $m$ required for this resonance instability to emerge can be approximated by 
	\begin{equation}
		\begin{gathered}
			m^{*} = 4N\dfrac{N-1}{N-2}\sqrt{N^2-4N+3}
		\end{gathered}
		\label{eq:parametric_location}
	\end{equation}
	which together with $\alpha^{*} = \arcsin\left(\dfrac{1}{N-2}\right)$ yield the lowest point of the resonance instability region in the $(\alpha,m)$ plane (the orange points in Fig.~\ref{main}).
}


\begin{figure}[h!]\center
	\includegraphics[width=8.2cm]{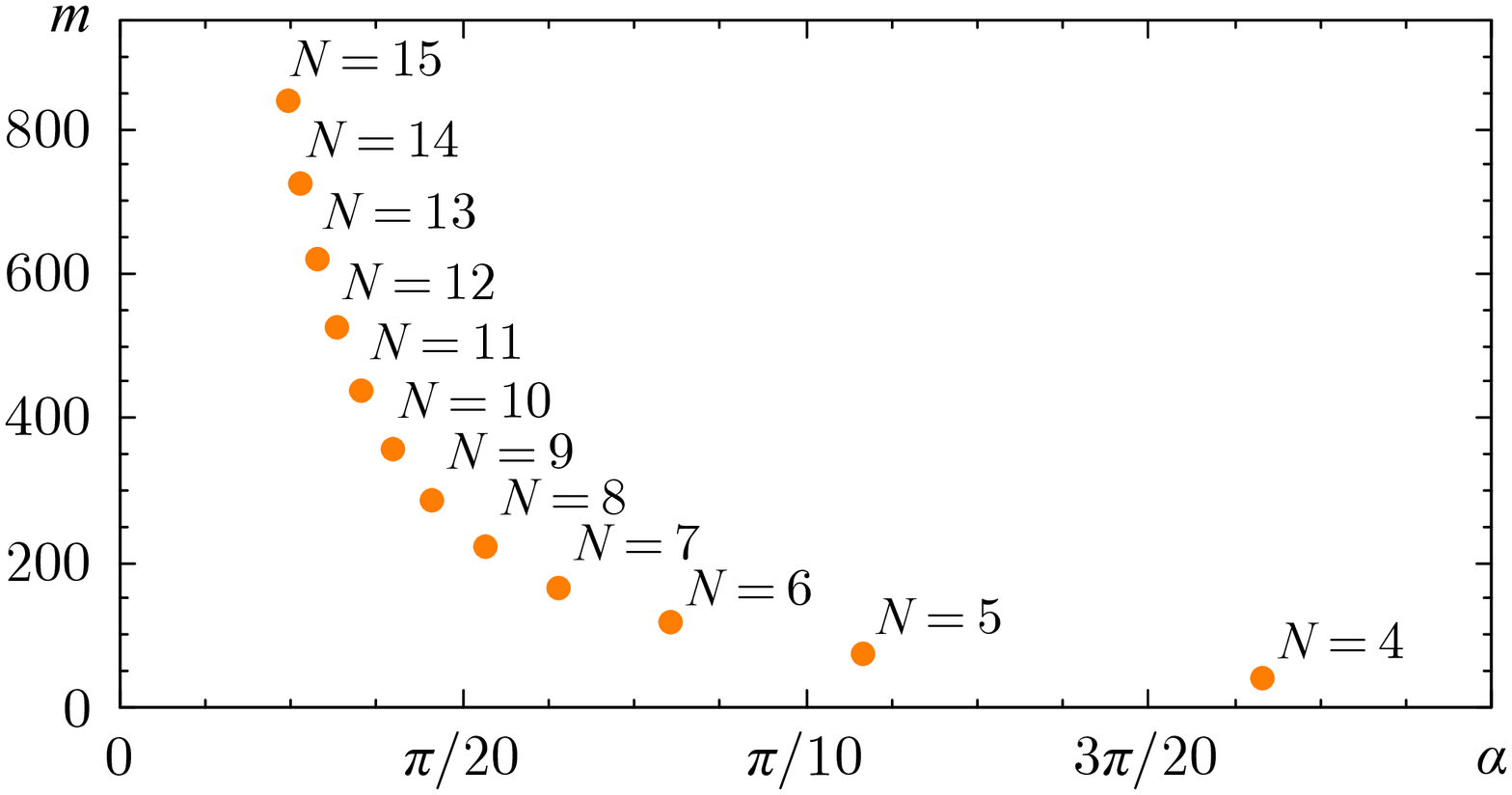}
	\caption{The lowest point of the parametric resonance region, $(\alpha^*,m^*),$ calculated via \eqref{eq:parametric_location}, as a function of network size $N.$ Note that increasing $N$ shifts the point
	to higher values of $m,$ thereby pushing the instability region in the upward direction and enlarging the overall stability region for the rotatory solitary state.}\label{lowest}
\end{figure}

\begin{figure}[h!]\center
	\includegraphics[width=8.5cm]{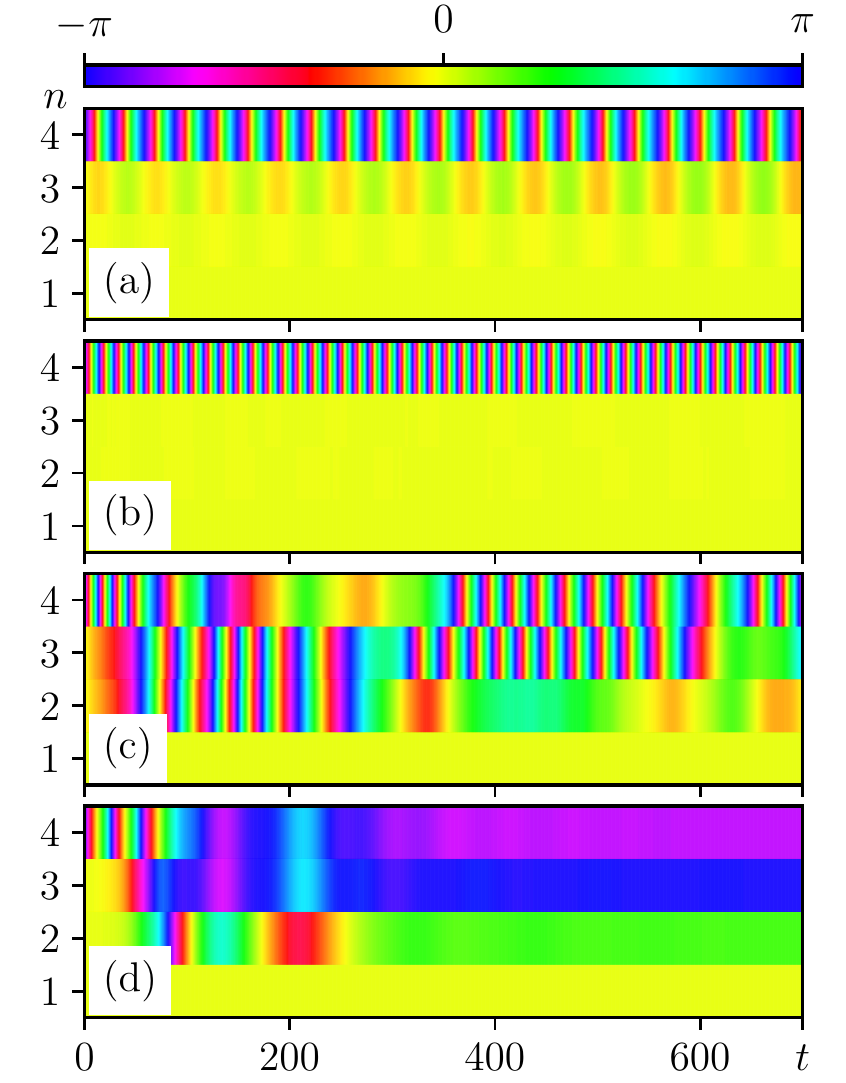}
	\caption{Illustration of Scenarios A and B for emergent instability of the rotatory solitary state in the four-oscillator network, corresponding to Fig.~\ref{main}(b). Time series of phase differences $\theta_n(t) - \theta_1(t),$ $n=1,...,4$ for varying $\alpha$ and fixed $m=60.$ Oscillator with index $n=1$ is used as a reference, while the solitary oscillator has index $n=4.$ The horizontal color bar uses the full range of 
	$\theta_n\in [-\pi,\pi].$ (a) Scenario B for $\alpha=0.45$ that belongs to the parametric resonance region in Fig.~\ref{main}(b): the rotatory solitary state is unstable and evolves into a periodic cluster regime. (b). Stable rotatory solitary state for $\alpha = \pi/3$. (c)  Scenario A for $\alpha=\pi/2 +0.01$: 
	the unstable solitary state turns  into a chaotic cluster regime. (d) Unstable solitary state becomes a generalized splay state for $\alpha=5\pi/6$. Initial conditions are chosen randomly as small perturbations of the solitary state. Parameter $\omega=1.$} \label{scenarios}
\end{figure}

\textit{Proof}. By virtue of \eqref{double}, we introduce a small parameter $\beta=\hat{\Omega}-\omega_x/2 \ll 1$ and 
seek a solution of variational equation \eqref{eq:Mathieu} in the form
\begin{equation}
	\xi\left(t\right)=a\left(t\right)\cos\frac{\omega_x t}{2}+b\left(t\right)\sin\frac{\omega_x t}{2},
	\label{eq:xi_ab1}
\end{equation}
where $a(t)$ and $b(t)$ are as yet undetermined slowly varying coefficients dependent on $\beta.$ To eliminate the arbitrariness in the introduction of two functions $a(t)$ and $b(t),$ we impose the constraint 
\begin{equation}
	\dot{a}\left(t\right)\cos\frac{\omega_{x}t}{2}+\dot{b}\left(t\right)\sin\frac{\omega_x t}{2}=0.
	\label{eq:xi_ab2}
\end{equation}
Substituting solution \eqref{eq:xi_ab1} into \eqref{eq:Mathieu} and taking into account \eqref{eq:xi_ab2},
we can separate the equations governing the evolutions of $a(t)$ and $b(t)$ as follows 
\begin{widetext}
	\begin{equation}
		\begin{aligned}
			&\dot{a}=-\frac{1}{m}\left(a\sin^2\frac{\omega_{x}t}{2}-\frac{b}{2}\sin\omega_{x}t\right)+\frac{2}{\omega_x}\left(
			\beta \left [ \hat{\Omega}+\frac{\omega_x}{2}\right ]
						+q\hat{\Omega}^2\cos\left(x\left(t\right)-\alpha\right)\right)\left(\frac{a}{2}\sin\omega_{x}t+b\sin^2\frac{\omega_{x}t}{2}\right), \\
			&\dot{b}=\frac{1}{m}\left(\frac{a}{2}\sin\omega_{x}t-b\cos^2\frac{\omega_{x}t}{2}\right)-\frac{2}{\omega_x}\left(
			\beta \left [ \hat{\Omega}+\frac{\omega_x}{2}\right ]
					+q\hat{\Omega}^2\cos\left(x\left(t\right)-\alpha\right)\right)\left(a\cos^2\frac{\omega_{x}t}{2}+\frac{b}{2}\sin\omega_{x}t\right).
		\end{aligned}
		\label{eq:ab}
	\end{equation}
\end{widetext}
We assume that the inertia and network size are sufficiently large such that $m \gg 1$ and $N \gg 1,$  making parameters $m^{-1} \ll 1$ and $q \ll 1$. Therefore,  the right-hand sides of \eqref{eq:ab} contain small parameters  $m^{-1}$, $q$ and $\beta,$ thereby allowing the application of the Van der Pol averaging method \cite{andronov2013theory}. To perform such averaging of the right-hand side of \eqref{eq:ab} over the period of $x(t)$, $T_x,$ in the first order of approximation, we take into account only the leading term in \eqref{approx} so that
$x\left(t\right)\approx\omega_{x}t.$ Under this condition, time averaging of \eqref{eq:ab} yields
\begin{equation*}
	\begin{gathered}
		\dot{a}=-\frac{a}{2}\!\left(\frac{1}{m}\!-\!\frac{q\hat{\Omega}^2}{\omega_x}\sin\alpha\right)\!+\!\frac{b}{\omega_x}\!\left(\beta\left(\omega_x+\beta\right)\!-\!\frac{q\hat{\Omega}^2}{2}\cos\alpha\right), \\
		\dot{b}=-\frac{a}{\omega_x}\!\left(\beta\left(\omega_x+\beta\right)\!+\!\frac{q\hat{\Omega}^2}{2}\cos\alpha\right)\!-\!\frac{b}{2}\!\left(\frac{1}{m}\!+\!\frac{q\hat{\Omega}^2}{\omega_x}\sin\alpha\right).
	\end{gathered}
\end{equation*}
Since only the first approximation is considered, we set $\beta^2=0$  and $\hat{\Omega}/\omega_x =1/2$ to obtain
\begin{equation}
	\begin{gathered}
		\begin{pmatrix} \dot{a} \\ \dot{b} \end{pmatrix} =
		\begin{pmatrix}
			-\dfrac{1}{2m}\!+\!\dfrac{q\hat{\Omega}}{4}\sin\alpha &
			\beta\!-\!\dfrac{q\hat{\Omega}}{4}\cos\alpha \\
			-\beta\!-\!\dfrac{q\hat{\Omega}}{4}\cos\alpha &
			-\dfrac{1}{2m}\!-\!\dfrac{q\hat{\Omega}}{4}\sin\alpha
		\end{pmatrix}
		\begin{pmatrix} a \\ b \end{pmatrix}.
	\end{gathered}\label{ab}
\end{equation}
The roots of the characteristic equation associated with linear system \eqref{ab} are
\begin{equation*}
	p_{1,2}=-\frac{1}{2m}\pm\sqrt{\left(\frac{q\hat{\Omega}}{4}\right)^2-\beta^2}.
\end{equation*}
Thus, system \eqref{ab} is unstable if
\begin{equation}
	\frac{1}{2m}<\sqrt{\left(\frac{q\hat{\Omega}}{4}\right)^2-\beta^2}.
	\label{eq:mathieu_instability}
\end{equation}
Under this condition, functions $a(t)$ and $b(t)$ grow without bounds, making solution \eqref{eq:xi_ab1} unstable and inducing parametric resonance instability in variational equation  \eqref{eq:Mathieu}. 
Expressing $\hat{\Omega}$,  $q$, $\beta$ via the original parameters of network \eqref{eq:mainSys} and using the first order approximation $\omega_x\approx\Omega_p=\frac{N-2}{N}\sin\alpha$ (cf. \eqref{approx}), we arrive at  approximate bound \eqref{resonance} that predicts the emergent instability via parametric resonance. This completes the proof of Part A.

The critical point of implicit function \eqref{resonance} corresponding to the lowest point of the instability region in the $(\alpha,m)$ plane could have been identified directly from \eqref{resonance}; however, these calculations are cumbersome and are not given here. Instead, we seek to find a locus of points inside the instability region, given by the condition $\beta=0$, i.e. $2\hat{\Omega}=\omega_x\approx\frac{N-2}{N}\sin\alpha.$ Expressing these constants via $\alpha$ and $m,$ we obtain 
\begin{equation}
	m_{locus}\left(\alpha\right)=4N\frac{N-1}{\left(N-2\right)^2}\frac{\cos\alpha}{\sin^2\alpha}.
	\label{eq:mathieu_mp}
\end{equation}

It follows from \eqref{eq:mathieu_instability} that the locus of points lies inside the parametric resonance instability region if $\frac{1}{2m}<\frac{\hat{\Omega} q}{4}$, i.e.
\begin{equation}
	m>4N\left(N-1\right)\cos\alpha.
	\label{eq:mathieu_mex}
\end{equation}
Making inequality \eqref{eq:mathieu_mex} an equality, solving it for $\cos \alpha,$ and using the obtained expression to replace the trigonometric terms in \eqref{eq:mathieu_mp}, we arrive at condition 		\eqref{eq:parametric_location}. Similarly, solving this equality together with \eqref{eq:mathieu_mp}, we obtain
$\alpha^{*}=\arcsin\left(\dfrac{1}{N-2}\right)$ which corresponds to $m^*.$ This completes the proof of Part B. $\Box$

Figure~\ref{main} compares our analytical bounds \eqref{resonance} and \eqref{eq:parametric_location}  with the numerically calculated  parametric resonance region (the red crescent-shaped region) and shows that the bounds predict the location of the instability region quite well, especially for $N \ge 4.$ The gap between 
two branches (dotted orange lines) of bound \eqref{resonance} for $N=3$ (Fig.~\ref{main}(a)) is due to a singularity of function \eqref{resonance} close to the cusp point of the instability region. Yet, the upper branch of bound \eqref{resonance} practically coincides with the upper border of the instability region even for the smallest network size, $N=3.$ Remarkably, our analytical bounds, consistent with the direct numerical simulations, indicate that the size of the parametric resonance region decreases when $N$ increases (Figs.~\ref{main}(a)-(d)). Figure~\ref{lowest} provides a more detailed description of the evolution of the parametric resonance region via the change of its lowest point as a function of $N.$
Thus, increasing the network size in the range $\alpha\in (0,\pi/2)$ where the coupling is attractive increases the stability region via (i) shrinking the parametric resonance region towards its eventual disappearance at large $N$ and (ii) making the existence boundary, homoclinic bifurcation curve \eqref{m1}, surround a larger region of the $(\alpha,m)$ parameter plane (see the evolution of the solid black curve in Figs.~\ref{main}(a)-(d)). Note that such an increase of $N$ makes the size of the synchronous
cluster, $N-1,$ larger, supposedly promoting complete synchronization in the network and attracting the solitary oscillator.  Counterintuitively, the above analysis shows that this not the case.

Figure~\ref{scenarios} illustrates how the rotatory solitary state loses its stability via Scenarios A and B
while passing the parametric resonance region and entering the repulsive coupling region in Fig.~\ref{main}(b) for a fixed $m.$ Note the double oscillation frequency of the solitary oscillator with index $n=4$ in Fig.~\ref{scenarios}(a) corresponding to the parametric resonance region. Also note in Fig.~\ref{scenarios}(d) that the strong repulsive coupling turns the solitary state into a fascinating dynamical pattern such as a generalized splay state \cite{berner2021generalized} in which the oscillators are frequency synchronized; yet,
possessing a vanishing order parameter.

\section{Resilience of rotatory  solitary states}
\begin{figure*}\center
	\includegraphics[width=1.7\columnwidth]{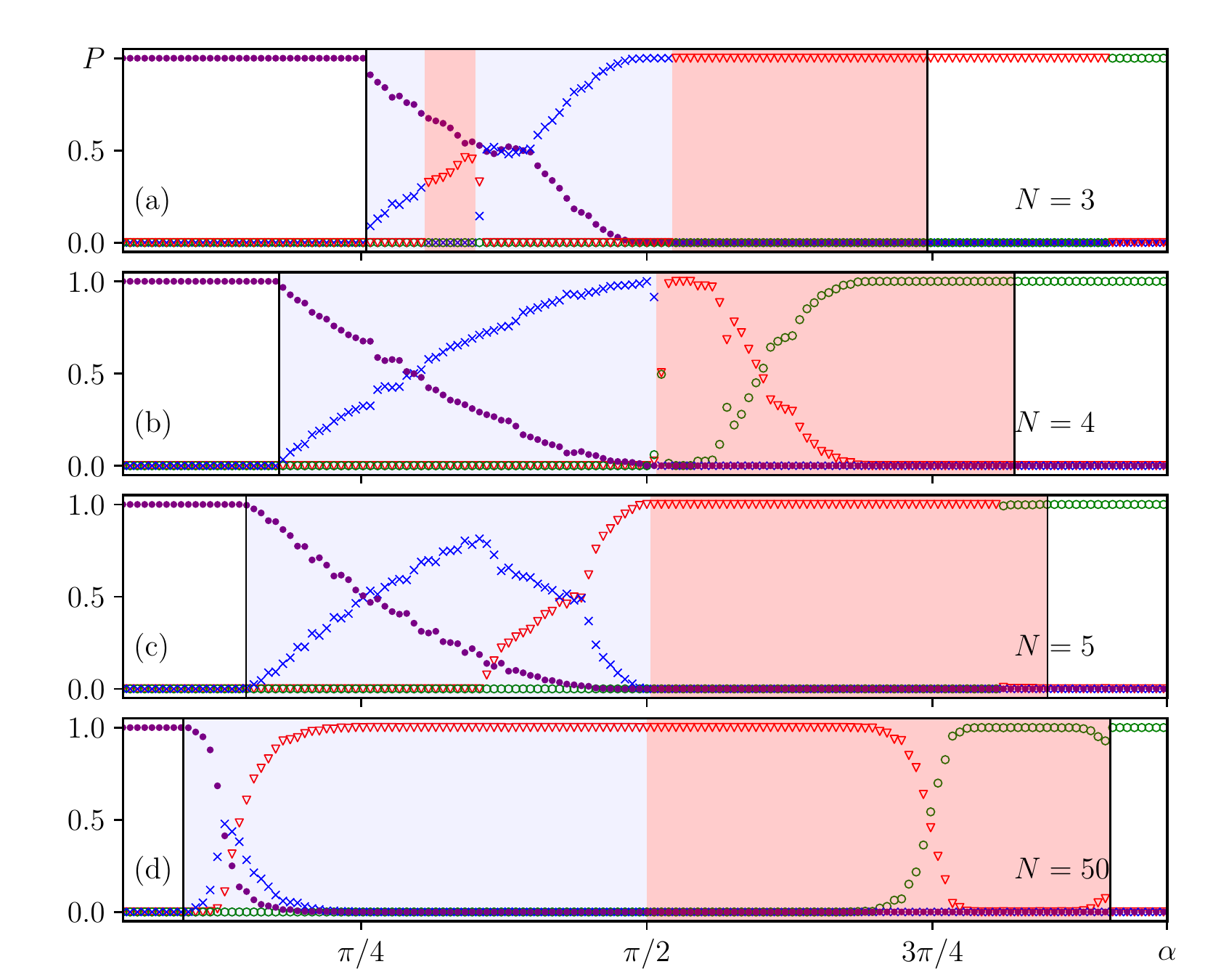}
	\caption{Probability of dynamical regimes' onset (blue crosses -- a solitary state, purple solid circles -- full synchronization, green empty circles -- a generalized splay state, red triangles -- other regimes). Random initial conditions $\theta_i(0)$ and $\dot{\theta}_i(0),$ $i=\overline{1,N}$ are evenly distributed within $[-\pi, \pi]$ and $[-10, 10],$ respectively.  Number of trials $1000.$
	(a). $N=3.$ (b). $N=4$. (c). $N=5$. (d). $N=50.$ Parameters $m=20$ and $\omega=1.$ The vertical black lines indicate the existence boundary of a solitary state. The light blue (light pink) region corresponds to the stability (instability) of a solitary state. Note the high probability of emergence of solitary states for $N=3$ and $N=4$ at $\alpha$ close to $\pi/2.$}\label{probability}
\end{figure*}

\begin{figure}[h!]\center
	\includegraphics[width=8.5cm]{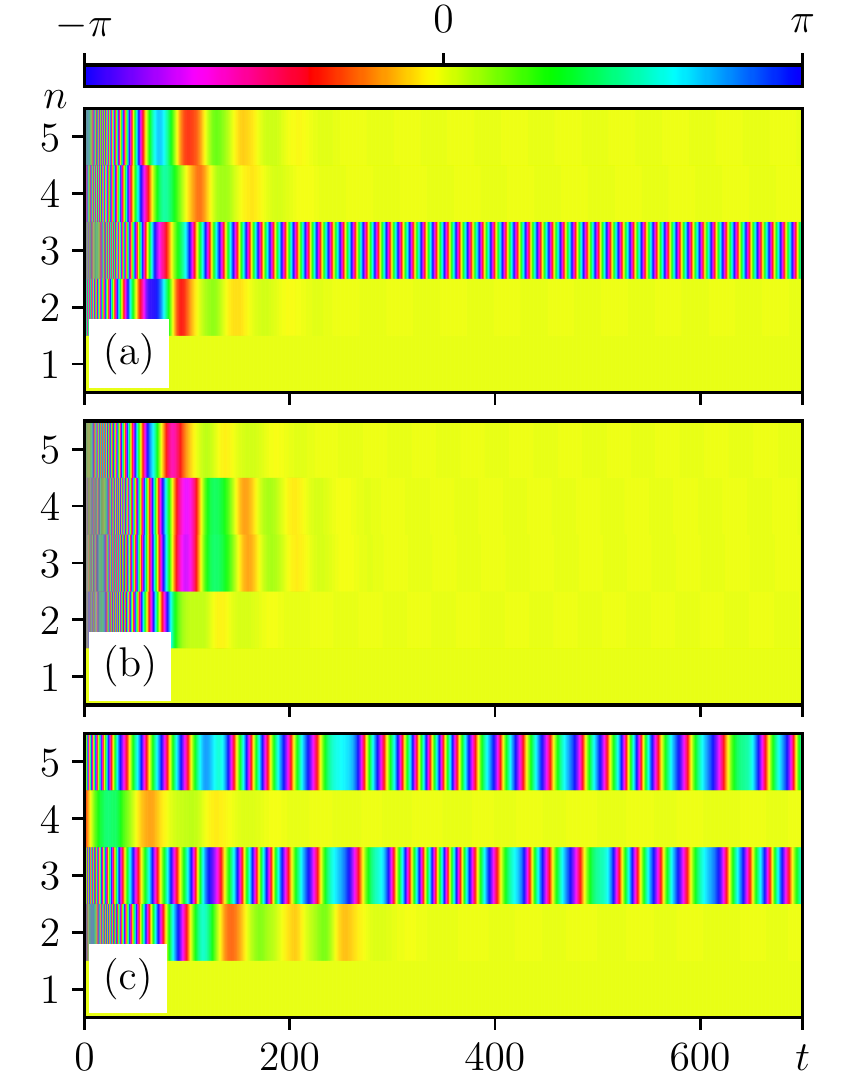}
	\caption{Co-existence of a stable solitary state (a), full synchronization (b), and a chaotic cluster regime (c) for $N = 5$, $m=80$, $\alpha = 1.2$, $\omega = 1$. The notations are as in Fig.~\ref{scenarios}. Random initial conditions are chosen as in Fig.~\ref{probability}.}\label{coexistence}
\end{figure}

The stability analysis performed in Sec.~\ref{stability} indicates that rotatory solitary states can be locally stable in a wide range of network parameters. However, these stable solitary states typically co-exist with other stable regimes, notably with full synchronization when the coupling is attractive for $\alpha \in (0,\pi/2).$
To test the resilience of stable rotatory solitary states to significant changes in the initial conditions and estimate the corresponding basins of attraction, we perform a numerical experiment reported in
Fig.~\ref{probability}. For each value of $\alpha \in (0,\pi),$ we perform 1000 numerical runs of the network from randomly chosen initial conditions (see the captions of Fig.~\ref{probability} for details) and count the number of times a solitary state emerges in the network. We also register the probabilities of the other outcomes which include full synchronization and generalized splay states. Our simulations indicate a high occurrence probability of solitary states in small networks in a range of $\alpha \in (\pi/3,\pi/2)$ where the emergence of solitary states is more probable than of full synchronization. Remarkably, 
solitary states become a dominant regime with probability $p$ close to 1 in a region of $\alpha$ around $\pi/2$ in the three- and four-oscillator networks (see two upper panels in Fig.~\ref{probability}). However, their emergence in this parameter region becomes improbable in the five-oscillator network
(third panel in Fig.~\ref{probability}). Figure~\ref{coexistence} displays a stable solitary state that co-exists with full synchronization and other regimes for the same value of network parameters.

\section{Conclusions}
Rotatory solitary states with one or more solitary oscillators can be viewed 
as particular examples of chimera states, called  ``weak chimeras''  \cite{maistrenko2017smallest}.  While a rigorous stability analysis of a fully developed chimera with a multi-oscillator incoherent state is typically out reach for finite-size networks, solitary states can offer a unique test bed for the development of   stability approaches to large chimeras. In this paper, we have developed such an approach and made significant progress in understanding stability properties of rotatory solitary states in finite size Kuramoto-Sakaguchi networks with inertia.

We have extended the previous work on the emergence of stable clusters in the second-order Kuramoto-Sakaguchi model \cite{belykh2016bistability} by deriving asymptotic stability conditions for a solitary state in which the phase-difference between the solitary oscillator and the remaining synchronous cluster changes periodically. In particular, we have analytically predicted two bifurcation scenarios by which such a solitary state 
can lose its stability. The first, main scenario is associated with the loss of stability near the transition from attractive to repulsive coupling when phase lag $\alpha$ exceeds $\pi/2.$ The unusual feature of this
well-known transition is that solitary states can remain stable and even become more resilient to perturbations in a range of weak repulsive coupling near  $\alpha=\pi/2,$ provided that the network size is small. The second, less common, scenario is associated with parametric resonance instability when the frequency of periodically varying phase difference is approximately twice the characteristic frequency of transversal perturbations of the solitary state. 
Through our asymptotic analysis, we have also revealed a fascinating property of solitary states to become stable in a larger region of parameters with an increase in network size. This happens despite the increase of synchronous cluster size that could supposedly strengthen the force aiming to bring the solitary oscillator back to the cluster.

We have performed  our asymptotic analysis under the assumption of large inertia and large network sizes; however, our approximate bounds proved to be quite accurate for intermediate values of inertia and small networks
with $N>3.$ The occurrence of stable rotatory solitary states in the second-order Kuramoto-Sakaguchi model was  previously studied numerically in \cite{maistrenko2017smallest,jaros2015chimera,jaros2018solitary}, and similar instability regions were reported (see Fig.~1 for $N=3$ in \cite{maistrenko2017smallest}). Our results provide analytical support to this numerical study and reveal the origins of emergent instability and their explicit dependence on network size $N$, inertia $m$, and phase lag $\alpha.$

Our analysis combined with the auxiliary system approach developed for identical oscillator networks in \cite{brister2020three} can be applied to solitary states with more than one solitary oscillator, representing ``stronger'' chimeras as well as to regular multi-cluster states. In such settings, the dynamics of phase differences are governed by a system of coupled pendulum-type equations \cite{brister2020three}. Yet, 
the approach from \cite{brister2020three} can characterize the multidimensional dynamics via a lower-dimensional auxiliary system while our asymptotic analysis can handle its stability.

Our approach is also applicable to solitary states in second-order Kuramoto networks with mixed attractive and repulsive connections. While the occurrence of solitary states in mixed first-order Kuramoto networks was previously studied in detail in \cite{maistrenko2014solitary,teichmann2019solitary}, the addition of inertia which increases the dimensionality of phase difference dynamics can induce unexpected synergistic effects. These effects can emerge from a combination of the two scenarios of instability and reverse the roles of couplings, similarly to mixed networks of excitation and inhibitory neurons \cite{belykh2015synergistic}. These problems are a subject of future study.

\section{Acknowledgment}
This work was supported by the Ministry of Science and Higher Education of the Russian Federation (Sec.~II, III and IV, Project No. 0729-2020-0036), the Russian Science Foundation (Sec.~V, Project No. 19-12-00367), the Scientific and Education Mathematical Center ``Mathematics for Future Technologies'' (Appendices A and B, Project No. 075-02-2020-1483/1) and the National Science Foundation (USA) under grants Nos. DMS-1909924 and CMMI-2009329 (to I.V.B.).

\section{Appendix A: Approximate solitary state solution}
To derive an approximate periodic solitary state solution for $x(t)$ under the assumption of large inertia, $m \gg 1,$ we shall follow the steps of the Lindstedt--Poincar\'e method \cite{drazin1992nonlinear}. Our goal is to introduce a new slow time and approximate the solution and the time scaling via asymptotic series, while avoiding secular terms.

Assume that $t=\sqrt{m}\varpi\tau$, where the unknown parameter $\varpi$ can be expanded as a power series of small parameter $\varepsilon=1/\sqrt{m}\ll 1$:
\begin{equation}
	\varpi=\sum_{j=0}^{\infty}\varepsilon^{j}\varpi_j.
	\label{eq:omega_expansion}
\end{equation}
Rescaling the time in equation \eqref{eq:detuning}, we obtain the equation for $x\left(\tau\right)$:
\begin{equation}
	\begin{gathered}
		x''+\varepsilon\varpi x'+\varpi^2\left(\frac{N-1}{N}\sin{\left(x+\alpha\right)}+\frac{1}{N}\sin{\left(x-\alpha\right)}\right)\\
		=\varpi^2\frac{N-2}{N}\sin{\alpha},
	\end{gathered}
	\label{eq:xr_tau_eq}
\end{equation}
where the prime notation is used to denote derivatives with respect to $\tau$. Without loss of generality, we disregard transients and consider the established limit cycle regime with 
initial condition $x\left(0\right)=0$ for $x\left(\tau\right)$. We seek solution $x\left(\tau\right)$ as the following expansion:
\begin{equation}
	x\left(\tau\right)=\tau+\sum_{j=0}^{\infty}\varepsilon^{j}x_j\left(\tau\right),
	\label{eq:xr_expansion}
\end{equation}
where $x_j$ are $2\pi$-periodic functions of $\tau$. Substituting~\eqref{eq:omega_expansion} and~\eqref{eq:xr_expansion} into~\eqref{eq:xr_tau_eq} and setting the terms of the same order of $\varepsilon$ to zero, we first obtain: 
\begin{equation}
	\begin{gathered}
		x_0''+\varpi_0^2\left(\frac{N-1}{N}\sin{\left(\tau+x_0+\alpha\right)}+\frac{1}{N}\sin{\left(\tau+x_0-\alpha\right)}\right)\\
		=\varpi_0^2\frac{N-2}{N}\sin{\alpha}.
	\end{gathered}
	\label{eq:xr_0_eq}
\end{equation}
The right-hand side of equality~\eqref{eq:xr_0_eq} contains secular term $\varpi_0^2\frac{N-2}{N}\sin{\alpha}$, which yields an unlimited growth of component $x_0\left(\tau\right)$ ($\sim\tau^2$) in asymptotic series~\eqref{eq:xr_expansion}, thereby contradicting the assumption of the stable limit cycle. The contradiction can be avoided by setting $\varpi_0=0$ which corresponds to trivial solution 
$x_0\left(\tau\right)=0$. It directly follows from $\varpi_0=0$ and $x_0\left(\tau\right)=0$ that $x_1\left(\tau\right)=0$ and for the second power of $\varepsilon,$ we obtain:
\begin{equation}
	\begin{gathered}
		x_2''+\varpi_1^2\left(\frac{N-1}{N}\sin\left(\tau+\alpha\right)+\frac{1}{N}\sin\left(\tau-\alpha\right)\right)\\
		=\varpi_1\left(\varpi_1\frac{N-2}{N}\sin\alpha-1\right).
	\end{gathered}
	\label{eq:xr_2_eq}
\end{equation}
Setting the secular term on the right-hand side of \eqref{eq:xr_2_eq} equal to zero, we obtain $\varpi_1=\frac{N}{\left(N-2\right)\sin\alpha}$. In this case, the nontrivial component $x_2(\tau)$ in the expansion~\eqref{eq:xr_expansion}
becomes 
\begin{equation} 
	x_2\left(\tau\right)=\varpi_1\left(\cos\tau+\varpi_1\cos\alpha\sin\tau-1\right).
	\label{eq:xr_2_sol}
\end{equation}
Similarly, we continue the power series expansion to find all necessary terms $\varpi_j$ and $x_j\left(\tau\right)$ in asymptotic series~\eqref{eq:omega_expansion} and~\eqref{eq:xr_expansion}.  These terms up to the eight order of $\varepsilon$ constituting the approximate solution \eqref{approx} are listed below and derived via symbolic calculations in MATHEMATICA v.~11:
\begin{widetext}
	\begin{align*}
		&\varepsilon^3\!:
		&&x_3''+2\varpi_2\left(\cos\tau+\varpi_1\cos\alpha\sin\tau\right)=\varpi_2\hspace{1.5mm}\Rightarrow\hspace{1.5mm}
		\varpi_2=0,\hspace{1.5mm}
		x_3\left(\tau\right)=0.\\
		&\varepsilon^4\!:
		&&x_4''+2\varpi_3\left(\cos\tau+\varpi_1\cos\alpha\sin\tau\right)+\varpi_1^3\cos\alpha\cos2\tau+\frac{\varpi_1^2}{2}\left(\varpi_1^2\cos^2\!\alpha-1\right)\sin2\tau=\varpi_3\hspace{1.5mm}\Rightarrow \\
		&&&\varpi_3=0,\hspace{1.5mm} x_4\left(\tau\right)=\frac{\varpi_1^3}{4}\cos\alpha\left(\cos2\tau-1\right)+\frac{\varpi_1^2}{8}\left(\varpi_1^2\cos^2\!\alpha-1\right)\sin2\tau.\\
		&\varepsilon^5\!:
		&&x_5''+2\varpi_4\left(\cos\tau+\varpi_1\cos\alpha\sin\tau\right)=\varpi_4\hspace{1.5mm}\Rightarrow\hspace{1.5mm}
		\varpi_4=0,\hspace{1.5mm}
		x_5\left(\tau\right)=0.\\
		&\varepsilon^6\!:
		&&x_6''+\left(\!2\varpi_5-\frac{\varpi_1^3}{16}\left(9\varpi_1^2\cos^2\!\alpha+13\right)\!\right)\cos\tau+\varpi_1\cos\alpha\left(\!2\varpi_5-\frac{\varpi_1^3}{16}\left(5\varpi_1^2\cos^2\!\alpha+9\right)\!\right)\sin\tau+\frac{\varpi_1^3}{4}\left(1-\varpi_1^2\cos^2\!\alpha\right)\cos2\tau \\
		&&&+\frac{\varpi_1^4}{2}\cos\alpha\sin2\tau+\frac{3}{16}\varpi_1^3\left(3\varpi_1^2\cos^2\!\alpha-1\right)\cos3\tau+\frac{3}{16}\varpi_1^4\cos\alpha\left(\varpi_1^2\cos^2\!\alpha-3\right)\sin3\tau=\varpi_5-\frac{\varpi_1^3}{2}\left(1+\varpi_1^2\cos^2\!\alpha\right)\Rightarrow \\
		&&&\varpi_5=\frac{\varpi_1^3}{2}\left(1+\varpi_1^2\cos^2\!\alpha\right), \\
		&&&x_6\left(\tau\right)=\frac{1}{24}\left(5\varpi_1^3-21\varpi_5\right)+\frac{1}{8}\left(7\varpi_5-2\varpi_1^3\right)\cos\tau+\frac{\varpi_1}{8}\left(11\varpi_5-2\varpi_1^3\right)\cos\alpha\sin\tau \\
		&&&+\frac{1}{8}\left(\varpi_1^3-\varpi_5\right)\cos2\tau+\frac{\varpi_1^4}{8}\cos\alpha\sin2\tau+\frac{1}{24}\left(3\varpi_5-2\varpi_1^3\right)\cos3\tau+\frac{\varpi_1}{24}\left(\varpi_5-2\varpi_1^3\right)\cos\alpha\sin3\tau. \\
		&\varepsilon^7\!:
		&&x_7''+2\varpi_6\left(\cos\tau+\varpi_1\cos\alpha\sin\tau\right)=\varpi_6\hspace{1.5mm}\Rightarrow\hspace{1.5mm}
		\varpi_6=0,\hspace{1.5mm}
		x_7\left(\tau\right)=0.\\
		&\varepsilon^8\!:
		&&x_8''+\left(2\varpi_7-\frac{\varpi_1^2}{16}\cos\alpha\left(2\varpi_1^3+3\varpi_5\right)\right)\cos\tau+\varpi_1\left(2\varpi_7\cos\alpha+\frac{1}{16}\left(2\varpi_1^3+3\varpi_5\right)\right)\sin\tau \\
		&&&+\frac{\varpi_1^2}{4}\cos\alpha\left(10\varpi_5-\varpi_1^3\right)\cos2\tau+\frac{\varpi_1}{4}\left(\varpi_1^3-11\varpi_5+12\frac{\varpi_5^2}{\varpi_1^3}\right)\sin2\tau+\frac{5}{16}\varpi_1^2\cos\alpha\left(2\varpi_1^3-\varpi_5\right)\cos3\tau \\
		&&&+\frac{5}{16}\varpi_1\left(3\varpi_5-2\varpi_1^3\right)\sin3\tau+\frac{\varpi_1^2}{2}\cos\alpha\left(\varpi_5-\varpi_1^3\right)\cos4\tau+\frac{\varpi_1}{4}\left(2\varpi_1^3-4\varpi_5+\frac{\varpi_5^2}{\varpi_1^3}\right)\sin4\tau=\varpi_7\hspace{1.5mm}\Rightarrow \\
		&&&\varpi_7=0,\\
		&&&x_8\left(\tau\right)=\frac{\varpi_1^2}{288}\cos\alpha\left(43\varpi_1^3-125\varpi_5\right)-\frac{\varpi_1^2}{16}\cos\alpha\left(2\varpi_1^3+3\varpi_5\right)\cos\tau+\frac{\varpi_1}{16}\left(2\varpi_1^3+3\varpi_5\right)\sin\tau \\
		&&&+\frac{\varpi_1^2}{16}\cos\alpha\left(10\varpi_5-\varpi_1^3\right)\cos2\tau+\frac{\varpi_1}{16}\left(\varpi_1^3-11\varpi_5+12\frac{\varpi_5^2}{\varpi_1^3}\right)\sin2\tau+\frac{5}{144}\varpi_1^2\cos\alpha\left(2\varpi_1^3-\varpi_5\right)\cos3\tau \\
		&&&+\frac{5}{144}\varpi_1\left(3\varpi_5-2\varpi_1^3\right)\sin3\tau+\frac{\varpi_1^2}{32}\cos\alpha\left(\varpi_5-\varpi_1^3\right)\cos4\tau+\frac{\varpi_1}{64}\left(2\varpi_1^3-4\varpi_5+\frac{\varpi_5^2}{\varpi_1^3}\right)\sin4\tau.
	\end{align*}
\end{widetext}

\section{Appendix B: Proof of Statement~1}
Our goal is to asymptotically approximate characteristic exponents $\Lambda$ and  $\hat{\Lambda},$  associated with multipliers $\mu_1$ and $\mu_2$ of equation \eqref{x}, respectively. We assume that $m \gg 1$ and $N \gg 1$ and seek solution $\xi\left(t\right)=e^{\Lambda t}\zeta\left(t\right),$ where  $\zeta\left(t\right)$ is a $T_x$-periodic function. Similarly to the asymptotic analysis performed in Appendix~A, we rescale time $t=\sqrt{m}\varpi\tau$, where $\varpi=\varepsilon\varpi_1+\varepsilon^5\varpi_5+o\left(\varepsilon^7\right)$ was calculated in Appendix~A,
and substitute $\xi\left(t\right)$ into \eqref{x}. We obtain the following equation
\begin{equation}
	\begin{gathered}
		\zeta''+\left(1+2\frac{\Lambda}{\varepsilon^2}\right)\varepsilon\varpi\zeta'+\\
		\varpi^2\left(\frac{\Lambda^2}{\varepsilon^2}+\Lambda+\dfrac{N-1}{N}\cos{\alpha}+\dfrac{1}{N}\cos{\left(x\left(\tau\right)-\alpha\right)}\right)\zeta=0,
	\end{gathered}
	\label{eq:zeta}
\end{equation}
where, as in Appendix~A, the prime notation denotes derivatives with respect to slow time $\tau$ and
$\varepsilon=1/\sqrt{m}$.

It is instructive to use the following substitution that helps to get rid of singular terms proportional to the reciprocal of $\varepsilon:$
\begin{equation}
	\Lambda=\varepsilon\sqrt{\lambda}-\varepsilon^2/2.
	\label{eq:lambda}
\end{equation}
Note that the other characteristic exponent, $\hat{\Lambda},$ can be calculated from Liouville's identity as follows:
\begin{equation}
	\hat{\Lambda}=-\varepsilon\sqrt{\lambda}-\varepsilon^2/2.
\end{equation}
However, $\hat{\Lambda}<0$ and therefore, does not induce instability and can be ignored.

Substituting \eqref{eq:lambda} into \eqref{eq:zeta}, we obtain
\begin{equation}
	\begin{gathered}
		\zeta''+2\sqrt{\lambda}\varpi\zeta'+\\
		\varpi^2\left(\lambda-\frac{\varepsilon^2}{4}+\dfrac{N-1}{N}\cos{\alpha}+\dfrac{1}{N}\cos{\left(x\left(t\right)-\alpha\right)}\right)\zeta=0.
	\end{gathered}
	\label{eq:zeta_l}
\end{equation}
Equation ~\eqref{eq:zeta_l} is suitable for constructing a regular asymptotic series for $\lambda$. To this end, we seek the $2\pi$-periodic solution $\zeta\left(\tau\right)$ and $\lambda$ in the form of the following expansions:
\begin{equation}
	\zeta\left(\tau\right)=\sum_{j=0}^{\infty}\varepsilon^j\zeta_j\left(\tau\right),\qquad
	\lambda=\sum_{j=0}^{\infty}\varepsilon^j\lambda_j,
\end{equation}
where $\zeta_j$ are $2\pi$-periodic functions of $\tau$. For the zeroth order of $\varepsilon$ we obtain   $\zeta_0''=0$ whose repeated integration gives $\zeta_0\left(\tau\right)=C,$ where $C$ is an arbitrary constant of integration. For definiteness, we set $\zeta_0\left(\tau\right)=1$. For simplicity, we will be setting all other arbitrary constants of integration obtained from calculations of high-order terms
$\zeta_k\left(\tau\right)$ equal to zero. Thus, for the first order of $\varepsilon$, we obtain  $\zeta_1''=0$ and $\zeta_1\left(\tau\right)=0$. Continuing the series expansion in $\varepsilon$, we arrive at the following sequence of equations and their solutions:
\begin{widetext}
	\begin{minipage}{0cm}
		\begin{align*}
			\varepsilon^2\!:\hspace{0.4cm}
			&\zeta_2''+\varpi_1^2\left(\lambda_0+\frac{N-1}{N}\cos\alpha\right)+\frac{\varpi_1^2}{N}\cos\!\left(\tau-\alpha\right)=0\hspace{1.5mm}\Rightarrow\hspace{1.5mm}\lambda_0=-\frac{N-1}{N}\cos\alpha,\hspace{1.5mm}\zeta_2\left(\tau\right)=\frac{\varpi_1^2}{N}\cos\!\left(\tau-\alpha\right).
		\end{align*}
	\end{minipage}
	
	\begin{minipage}{0cm}
		\begin{align*}
			\varepsilon^3\!:\hspace{0.4cm}
			&\zeta_3''+\varpi_1^2\lambda_1+2\sqrt{\lambda_0}\varpi_1\zeta_2'=0\hspace{1.5mm}\Rightarrow\hspace{1.5mm}\lambda_1=0,\hspace{1.5mm}\zeta_3\left(\tau\right)=-2\frac{\varpi_1^3}{N}\sqrt{\lambda_0}\sin\!\left(\tau-\alpha\right).
		\end{align*}
	\end{minipage}
	
	\begin{minipage}{0cm}
		\begin{align*}
			\varepsilon^4\!:\hspace{0.4cm}
			&\zeta_4''+\varpi_1^2\left(\lambda_2-\frac{1}{4}-\frac{1}{N}\sin\!\left(\tau-\alpha\right)x_2\right)+2 \sqrt{\lambda_0}\varpi_1\zeta_3'+\frac{\varpi_1^2}{N}\cos\!\left(\tau-\alpha\right)\zeta_2=0\hspace{1.5mm}\Rightarrow \\
			&\lambda_2=\frac{\varpi_1^2}{8N^2}\left(\left(N-2\right)^2-\left(N^2-8N+8\right)\cos2\alpha\right), \\
			&\zeta_4\left(\tau\right)=\frac{\varpi_1^4}{2N^2}\Big(3\left(N-1\right)\cos\tau+2\cos\alpha\cos\!\left(\tau-\alpha\right)+\left(5N-7\right)\cos\!\left(\tau-2\alpha\right)+\frac{N-1}{4}\cos2\tau+\frac{1}{2}\cos\!\left(2\tau-2\alpha\right)\Big).
		\end{align*}
	\end{minipage}
	
	\begin{minipage}{0cm}
		\begin{align*}
			\varepsilon^5\!:\hspace{0.4cm}
			&\zeta_5''+\varpi_1^2\lambda_3+\varpi_1\left(\frac{\lambda_2}{\sqrt{\lambda_0}}\zeta_2'+2\sqrt{\lambda_0}\zeta_4'\right)+\frac{\varpi_1^2}{N}\cos\!\left(\tau-\alpha\right)\zeta_3=0\hspace{1.5mm}\Rightarrow\hspace{1.5mm}\lambda_3=0, \\
			&\zeta_5\left(\tau\right)=\frac{\varpi_1^5}{16N^3\sqrt{\lambda_0}}\Big(\left(25N^2-48N+24\right)\sin\!\left(\tau+\alpha\right)+2\left(31N^2-60N+28\right)\sin\!\left(\tau-\alpha\right) \\
			&\hspace{1cm}+\left(41N^2-96N+56\right)\sin\!\left(\tau-3\alpha\right)+2\left(N-1\right)\cos\alpha\left(\left(N-1\right)\sin2\tau+4\sin\!\left(2\tau-2\alpha\right)\right)\Big).
		\end{align*}
	\end{minipage}
	
	\begin{minipage}{0cm}
		\begin{align*}
			\varepsilon^6\!:\hspace{0.4cm}
			&\zeta_6''+\varpi_1^2\left(\lambda_4-\frac{1}{2N}\cos\!\left(\tau-\alpha\right)x_2^2-\frac{1}{N}\sin\!\left(\tau-\alpha\right)x_4\right)+\varpi_1\left(\frac{\lambda_2}{\sqrt{\lambda_0}}\zeta_3'+2\sqrt{\lambda_0}\zeta_5'\right) \\
			&\hspace{3cm}+\varpi_1^2\left(\lambda_2-\frac{1}{4}-\frac{1}{N}\sin\!\left(\tau-\alpha\right)x_2\right)\zeta_2+\frac{\varpi_1^2}{N}\cos\!\left(\tau-\alpha\right)\left(\zeta_4+2\frac{\varpi_5}{\varpi_1}\right)=0\hspace{1.5mm}\Rightarrow \\
			&\lambda_4=\frac{\varpi_1^4}{2N^3}\left(N-1\right)\left(\left(N-2\right)\cos2\alpha-N-2\right)\cos\alpha,\\
			&\zeta_6\left(\tau\right)=\frac{\varpi_1^6}{16N^3}\Big(\left(N-1\right)^2\cos\!\left(\tau+3\alpha\right)+\left(53N^2-99N+53\right)\cos\!\left(\tau+\alpha\right)+\left(128N^2-232N+103\right)\cos\!\left(\tau-\alpha\right) \\
			&\hspace{1.5cm}+\left(87N^2-206N+124\right)\cos\!\left(\tau-3\alpha\right)+\left(N-1\right)\cos\!\left(2\tau+\alpha\right)+\left(2N^2+1\right)\cos\!\left(2\tau-\alpha\right) \\
			&\hspace{1.5cm}+\left(11N-14\right)\cos\!\left(2\tau-3\alpha\right)+\frac{\left(N-1\right)^2}{3}\cos\!\left(3\tau+\alpha\right)+\frac{11}{9}\left(N-1\right)\cos\!\left(3\tau-\alpha\right)+\cos\!\left(3\tau-3\alpha\right)\Big).
		\end{align*}
	\end{minipage}
	
	\begin{minipage}{0cm}
		\begin{align*}
			\varepsilon^7\!:\hspace{0.4cm}
			&\zeta_7''+\varpi_1^2\lambda_5+\varpi_1\left(\!\left(-\frac{\lambda_2^2}{4\lambda_0^{3/2}}+\frac{\lambda_4}{\sqrt{\lambda_0}}+2\sqrt{\lambda_0}\frac{\varpi_5}{\varpi_1}\right)\zeta_2'+\frac{\lambda_2}{\sqrt{\lambda_0}}\zeta_4'+2\sqrt{\lambda_0}\zeta_6'\right) \\
			&\hspace{3cm}+\varpi_1^2\left(\lambda_2-\frac{1}{4}-\frac{1}{N}\sin\!\left(\tau-\alpha\right)x_2\right)\zeta_3+\frac{\varpi_1^2}{N}\cos\!\left(\tau-\alpha\right)\zeta_5=0\hspace{1.5mm}\Rightarrow\hspace{1.5mm}\lambda_5=0, \\
			&\zeta_7\left(\tau\right)=-\frac{\varpi_1^7}{16N^5\sqrt{\lambda_0}}\Bigg(\frac{1}{\lambda_0}\Bigg(\frac{\left(N-1\right)^4}{2}\sin\!\left(\tau+5\alpha\right) \\
			&\hspace{3cm}+\left(\frac{1807}{64}N^4-\frac{449}{4}N^3+\frac{341}{2}N^2-\frac{467}{4}N+\frac{121}{4}\right)\sin\!\left(\tau+3\alpha\right) \\
			&\hspace{3cm}+\left(\frac{1953}{16}N^4-\frac{1899}{4}N^3+\frac{2831}{4}N^2-\frac{961}{2}N+\frac{251}{2}\right)\sin\!\left(\tau+\alpha\right) \\
			&\hspace{3cm}+\left(\frac{6525}{32}N^4-\frac{3197}{4}N^3+\frac{4751}{4}N^2-\frac{3171}{4}N+\frac{801}{4}\right)\sin\!\left(\tau-\alpha\right) \\
			&\hspace{3cm}+\left(\frac{2473}{16}N^4-\frac{2549}{4}N^3+\frac{3981}{4}N^2-697N+\frac{369}{2}\right)\sin\!\left(\tau-3\alpha\right) \\
			&\hspace{3cm}+\left(\frac{2863}{64}N^4-\frac{399}{2}N^3+\frac{1343}{4}N^2-252N+71\right)\sin\!\left(\tau-5\alpha\right)\!\Bigg) \\
			&\hspace{3cm}-\frac{N}{16}\Big(N^2\left(N-1\right)\sin\!\left(2\tau+2\alpha\right)+2\left(7N^3+19N^2-44N+20\right)\sin2\tau \\
			&\hspace{3cm}+\left(17N^3+183N^2-400N+192\right)\sin\!\left(2\tau-2\alpha\right)+4\left(47N^2-114N+68\right)\sin\!\left(2\tau-4\alpha\right)\Big) \\
			&\hspace{3cm}+\lambda_0\frac{N^2}{27}\left(6\left(N^2-2N-8\right)\sin\!\left(3\tau+\alpha\right)+\left(49N-49+108\cos2\alpha\right)\sin\!\left(3\tau-\alpha\right)\right)\!\Bigg).
		\end{align*}
	\end{minipage}
	
	\begin{minipage}{0cm}
		\begin{align*}
			\varepsilon^8\!:\hspace{0.4cm}
			&\zeta_8''+\varpi_1^2\left(\lambda_6+\frac{1}{6N}\sin\!\left(\tau-\alpha\right)x_2^3-\frac{1}{N}\cos\!\left(\tau-\alpha\right)x_2 x_4-\frac{1}{N}\sin\!\left(\tau-\alpha\right)x_6\right) \\
			&\hspace{2.5cm}+\varpi_1\left(\!\left(-\frac{\lambda_2^2}{4\lambda_0^{3/2}}+\frac{\lambda_4}{\sqrt{\lambda_0}}+2\sqrt{\lambda_0}\frac{\varpi_5}{\varpi_1}\right)\zeta_3'+\frac{\lambda_2}{\sqrt{\lambda_0}}\zeta_5'+2\sqrt{\lambda_0}\zeta_7'\right) \\
			&\hspace{2.5cm}+\varpi_1^2\bigg(\!\left(\lambda_4-\frac{1}{2N}\cos\!\left(\tau-\alpha\right)x_2^2-\frac{1}{N}\sin\!\left(\tau-\alpha\right)x_4+\frac{2}{N}\frac{\varpi_5}{\varpi_1}\cos\!\left(\tau-\alpha\right)\right)\zeta_2 \\
			&\hspace{2.5cm}+\left(\lambda_2-\frac{1}{4}-\frac{1}{N}\sin\!\left(\tau-\alpha\right)x_2\right)\left(\zeta_4+2\frac{\varpi_5}{\varpi_1}\right)+\frac{1}{N}\cos\!\left(\tau-\alpha\right)\xi_6\bigg)=0\hspace{1.5mm}\Rightarrow \\
			&\lambda_6=\frac{\varpi_1^6}{32N^4}\left(N-1\right)\left(4 \left(N^2-2N+2\right)\cos4\alpha+2\left(N^2-58N+70\right)\cos2\alpha+4N^2-133N+133\right).
		\end{align*}
	\end{minipage}
	
	Collecting the terms, we arrive at the final expression:
	\begin{equation}
		\begin{gathered}
			\lambda=-\frac{N-1}{N}\cos\alpha+\frac{\varpi_1^2}{8mN^2}\left(\left(N-2\right)^2-\left(N^2-8N+8\right)\cos2\alpha\right) \\
			+\frac{\varpi_1^4}{2m^2N^3}\left(N-1\right)\left(\left(N-2\right)\cos2\alpha-N-2\right)\cos\alpha \\
			+\frac{\varpi_1^6}{32m^3N^4}\left(N-1\right)\left(4\left(N^2-2N+2\right)\cos4\alpha+2\left(N^2-58N+70\right)\cos2\alpha+4N^2-133N+133\right)+o\left(m^{-3}\right).
		\end{gathered}
		\label{eq:sol_lambda}
	\end{equation}
\end{widetext}
Recall that we seek to find the condition under which the trivial solution of variational equation \eqref{x} undergoes a bifurcation associated with multiplier $\mu_1=+1$ and, therefore, with characteristic exponent  $\Lambda=0.$ Thus, setting $\Lambda=0$ in equation \eqref{eq:lambda}, replacing $\lambda$ with \eqref{eq:sol_lambda} and solving for $\alpha,$ we derive the bound of Statement~1:
\begin{equation}
	\alpha_c=\frac{\pi}{2}+\frac{N}{2m\left(N-2\right)^2}-\frac{N^3\left(18N^2-27N+13\right)}{96m^3\left(N-2\right)^6}+o\left(m^{-3}\right).\label{border}
\end{equation}
To support the claim made in Remark~1, we analyze the sign of $\Lambda$  near bifurcation curve \eqref{border}. To do so, we fix $m$ and consider $\Lambda$ as a function of $\alpha.$ If follows from 
\eqref{eq:lambda} that $\dfrac{d \Lambda}{d \alpha} \left(\alpha_c\right)=\dfrac{d\lambda}{d \alpha}\left(\alpha_c\right)$ since $\varepsilon$ is treated as a constant. This yields
\begin{equation}
\dfrac{d \Lambda}{d \alpha} \left(\alpha_c\right)=\frac{N-1}{N}+\frac{N\left(N-1\right)\left(8N+3\right)}{8m^2(N-2)^4}+o\left(m^{-3}\right)>0.
\end{equation}
Thus, characteristic exponent $\Lambda$ increases near the bifurcation curve and becomes zero at the curve. Therefore, $\Lambda<0$ ( $\Lambda>0$) for $\alpha<\alpha_c$ ( $\alpha>\alpha_c$) in the  vicinity of $\alpha_c.$ $\Box$

\bibliography{references_solitary_states}

\end{document}